\documentclass[12pt, a4paper]{article}
\usepackage{cmap}
\usepackage{cite}
\usepackage[T2A]{fontenc}			
\usepackage[utf8]{inputenc}			
\usepackage[ukrainian, english]{babel}	
\usepackage[colorlinks,hypertexnames=false]{hyperref}
\usepackage{cancel}
\usepackage{graphicx}  
\usepackage{wrapfig} 
\usepackage[justification=raggedright,singlelinecheck=false]{caption}

\usepackage{amssymb,amsfonts,amsmath,amsthm} 
\usepackage{icomma} 
\usepackage{textcomp} 
\usepackage{etoolbox}

\usepackage[usenames]{color} 
\usepackage{colortbl}
\usepackage[dvipsnames]{xcolor}

\usepackage{array,tabularx,tabulary,booktabs} 
\usepackage{longtable} 
\usepackage{multirow} 
\usepackage{multicol} 

\usepackage{bm} 
\usepackage{amssymb}
\usepackage{slashed}
\usepackage{mathtools}
\renewcommand{\phi}{\varphi}

\counterwithin{equation}{section}

\usepackage[justification=centering]{caption}
\usepackage{indentfirst}
\usepackage[titletoc, title]{appendix}

\begin{document}
\selectlanguage{english}
\title{Constraints on the parameters of the neutrino extension of the Standard Model}
\author{Volodymyr Gorkavenko${}^1$, Oleksandr Khasai${}^2$, Oleg Ruchayskiy${}^3$,\\ Mariia Tsarenkova${}^1$ \vspace{1em}\\
${}^1$ \it \small Faculty of Physics, Taras Shevchenko National University of Kyiv,\\ 
\it \small 64, Volodymyrs'ka str., Kyiv 01601, Ukraine,\\
${}^2$ \it \small Bogolyubov Institute for Theoretical Physics, National Academy of Sciences of Ukraine,\\
\it \small 14-b Metrolohichna str., Kyiv 03143, Ukraine,\\
${}^3$ \it \small Niels Bohr Institute, University of Copenhagen, \\
\it \small Blegdamsvej 17, DK-2010, Copenhagen, Denmark}
\date{}

\maketitle
\setcounter{equation}{0}
\setcounter{page}{1}%

\begin{abstract}
  Heavy neutral leptons (HNLs) are hypothetical particles proposed as a potential explanation for neutrino oscillations and the generation of the baryon asymmetry in the Universe. This paper focuses on HNLs with masses significantly above the electroweak scale.
  It is challenging to test for the presence of such particles directly.
  However, they leave behind effective interactions of Standard Model particles, leading in particular to charged lepton flavor violation (cLFV) processes. 
  Non-observation of cLFV processes puts therefore constraints on the parameters of the HNLs.
  In this paper, we find the relations between the effective operators in the realistic case when neutrino masses are non-zero and the HNLs are non-degenerate. This allows us to strengthen the existing cLFV constraints on the tau-flavors from much
stronger constraints on muon and electron flavors. 
  We also link the baryon asymmetry of the Universe to the same higher-dimensional effective operators, providing complementary bounds on these parameters. 
        
    \end{abstract}
    \selectlanguage{english}

\section{Introduction}\label{sec:intro}
The Standard Model of particle physics (SM) is a precise theory that accurately describes interactions involving elementary particles, specifically the electroweak and strong interactions \cite{Cottingham:2007zz}. It has been extensively tested in accelerator experiments up to an energy of approximately 15 TeV. It remains consistent up to a very high energy scale, potentially up to the Planck scale. However, the SM fails to explain certain phenomena such as the masses of neutrinos (see, e.g., \cite{Bilenky:1987ty,Strumia:2006db,deSalas:2017kay}), dark matter (see, e.g., \cite{Peebles:2013hla,Lukovic:2014vma,Bertone:2016nfn}), and the baryon asymmetry in the Universe (see, e.g., \cite{Steigman:1976ev,Riotto:1999yt,Canetti:2012zc}). Therefore, the SM is an incomplete theory and needs to be expanded. It is necessary to suggest the existence of "hidden" sectors containing particles of new physics to solve these problems.

The mentioned problems of the SM can potentially be solved by introducing new particles of different mass scales (ranging from heavy particles with masses up to the GUT scale to light particles with masses in the sub-eV range). It cannot be ruled out that there may also be new particles that are not directly related to a solution to the SM problems. 
The properties of these new particles, whether they are scalars, pseudoscalars, vectors, or fermions, remain unknown and need to be tested experimentally \cite{Beacham:2020}. Theoretical considerations suggest three possible choices for the renormalized interaction between the new particles and SM particles, referred to as portals: scalar \cite{Patt:2006fw,Bezrukov:2009yw,Boiarska:2019jym}, vector \cite{Okun:1982xi,Holdom:1985ag,Langacker:2008yv}, and fermion, or heavy neutral leptons (HNL) portals \cite{Bondarenko:2018ptm,Boyarsky:2018tvu}.  Other high-dimensional operators, such as the pseudoscalar (axion\-like) particle portal \cite{Peccei:1977hh, Weinberg:1977ma, Wilczek:1977pj,Choi:2020rgn} and portal with the Chern-Simons-like interaction \cite{Antoniadis:2009ze,Dror:2017nsg,Borysenkova:2021ydf} may also be of interest, see, e.g., \cite{Alekhin:2016,Curtin:2018mvb}.

This paper focuses on the extension of the SM with right-handed (RH) neutrinos, also known as sterile neutrinos, which have feeble interactions with SM particles. 
The addition of one sterile neutrino to the SM leads to a situation where only one active neutrino is massive, but two others are massless. This contradicts data on active neutrino oscillations.  The addition of two sterile neutrinos to the SM leads to the situation when two active neutrinos have different masses and one neutrino is massless, which is compatible with data of active neutrino oscillations (the case of normal and inverted hierarchy). 
The Neutrino Minimal Standard Model ($\nu$MSM), proposed in \cite{Asaka:2005,Asaka_2:2005}, extends the SM with three RH neutrinos $N_I$ ($I = 1,2,3$):
\begin{equation}\label{g1}
    \delta \mathcal{L} = i \bar N_I \partial_\mu\gamma^\mu N_I - F_{\alpha I} \bar L_\alpha \tilde\Phi N_I  - \frac{M_I}2 \bar N^c_I N_I + h.c.
\end{equation}
Here, $\alpha $ represents different flavors of leptons ($e, \mu, \tau$),  $\Phi$ and $L_\alpha$  are the Higgs and lepton doublets, respectively, $\tilde \Phi=i\tau_2 \Phi^*$, $F_{\alpha I}$ are elements of the Yukawa matrix, $M_I$ are Majorana masses of sterile neutrinos. 
It was shown that this model, with appropriate choices of its 18 new parameters (3 masses of active neutrinos, 3 masses of sterile neutrinos, 6 mixing angles, and 6 CP-violating phases), can simultaneously solve problems of neutrino oscillations, baryon asymmetry of the Universe, and dark matter. The lightest sterile neutrino in the $\nu$MSM is a particle of dark matter with a mass
of order $\sim\!10$ keV and a lifetime of order of the age of the Universe. Two other sterile neutrinos are heavy particles with almost equal masses. They
 provide generation of baryon asymmetry via the mechanism of leptogenesis \cite{Akhmedov:1998qx,Buchmuller:2004nz,Pilaftsis:2005rv,Davidson:2008bu,Pilaftsis:2009pk,Shaposhnikov:2009zzb,Bodeker:2020ghk} in the Universe and these heavy particles decay before the Big Bang nucleosynthesis.  Obtained in \cite{Asaka:2005} results for baryogenesis were revised for the case of 2 \cite{Klaric:2020phc,Klaric:2021cpi} and 3 \cite{Drewes:2021nqr} RH neutrinos. It should be noted that the condition of almost equal masses of the RH neutrinos for baryogenesis is not necessary, see, e.g., \cite{Drewes:2012ma}.

At energies well below the masses of RH neutrinos,  ($E\ll M_I$) their influence can be described in the framework of effective field theory of the Standard Model (SMEFT) \cite{Buchmuller:1985jz,Georgi:1991ch,Georgi:1992xg,Grzadkowski:2010es,Henning:2016lyp,Brivio:2017vri,DeVries:2020jbs}.  In this approach, at a low-energy scale $\mu < M_I$, the interaction terms involving SM particles and RH neutrinos can be replaced by a set of local effective operators $Q_i(\phi_{SM})$, which contain only the SM fields $\phi_{SM}$:
\begin{equation}
{\cal L}_{{\rm{EFT}}}\left( \phi  \right) = {{\cal L}_\phi }\left( \phi  \right) + \sum\limits_i {{c_i}\left(\mu\right){Q_i}\left( \phi_{SM}  \right)}, \label{eqn:LEFT}
\end{equation}
where ${Q_i}\left( \phi_{SM}  \right)$ 
are effective operators and $c_i\left(\mu\right)$ are Wilson coefficients at the energy scale $\mu$.  By expanding the tree-level heavy neutrino propagator  in a power series in $1/M_I$ \cite{Broncano:2002rw,Coy:2018bxr} one gets effective operators 
of specific interests for  lepton flavor violating (LFV) processes are the  following dimension-6 operators
\begin{equation}
    {\cal L}_{{\rm{EFT}}}^6=\frac{S_{\alpha\beta}}4 \left( Q_{Hl,{\alpha\beta}}^{(1)} -  Q_{Hl,{\alpha\beta}}^{(3)} \right),
\end{equation}
 where 
\begin{align}
    & Q_{Hl,{\alpha\beta}}^{(1)}=(\overline{l_{L\alpha}} \gamma_\mu l_{L\beta})(\Phi^\dagger i \overleftrightarrow{D^\mu} \Phi),\\
    & Q_{Hl,{\alpha\beta}}^{(3)}=(\overline{l_{L\alpha}} \gamma_\mu \sigma^A l_{L\beta})(\Phi^\dagger i \overleftrightarrow{D^\mu} \sigma^A \Phi)
\end{align}
are operators from the Warsaw basis \cite{Grzadkowski:2010es}  that generate charged lepton flavor violation (cLFV) processes and $S_{\alpha\beta}$ is defined as 
\begin{equation}
    S_{\alpha\beta} \equiv (F M^{-1\ast}M^{-1} F^\dagger)_{\alpha\beta} = \sum_I S^I_{\alpha\beta} = \sum_I F_{\alpha I}F^\dagger_{I \beta} M_I^{-2}.\label{Sab}
\end{equation}

To get corresponding operators at a lower energy scale (electroweak scale or mass of leptons), one has to apply renormalization group equations and find Wilson coefficients, see \cite{Broncano:2003fq,Jenkins:2013zja,Jenkins:2013wua,Alonso:2013hga,Ohlsson:2022hfl,Wang:2023bdw,Zhang:2023kvw,Zhang:2023ndw}. 
When solving the equations, the requirement is imposed that the one-light-particle irreducible diagrams (or effective action) computed from $\mathcal{L}_\text{SMEFT}$ and  computed from full theory agree at scale $\mu=M_I$. This is known as the matching criterion, see  \cite{Georgi:1991ch,Georgi:1992xg}. 
It turns out that in the one-loop approximation, 
these operators will be running and mixed into other operators. This gives rise to another characteristic of the process, see details in \cite{Coy:2018bxr},
\begin{equation}
    R_{\alpha\beta} = \sum_I R^I_{\alpha\beta} = \sum_I S^I_{\alpha\beta}\ln\frac{M_I}{M_W} = \sum_I F_{\alpha I}F^\dagger_{I \beta} M_I^{-2} \ln\frac{M_I}{M_W}.\label{Rab}
\end{equation}

In any case, because of the very large lifetime of the lightest sterile neutrino in $\nu$MSM (dark matter candidate), its coupling to SM particles is, obviously, significantly less than the couplings of the two other sterile neutrinos to SM particles.
Therefore, for experimental search at collider experiments, it is useful to employ a model with only two heavy sterile neutrinos, see e.g. \cite{Broncano:2003fq,Zhang:2021jdf,Coy:2021hyr,Du:2022vso}.


The observable reactions are sensitive to the parameters $S_{\alpha\beta}$ and $R_{\alpha\beta}$. For example, 
$Z$-boson flavor-conserving and flavor-violating decays are given by
\begin{align}
&\Gamma(Z \to \ell_\alpha^+ \ell_\alpha^-) \simeq \Gamma(Z \to \ell_\alpha^+ \ell_\alpha^-)^{\text{SM}}  \left[1 + \frac{v^2}{4} \frac{1 - 2 s_w^2 -4 s_w^4}{(1 - 2 s_w^2)(1-4s_w^2+8s_w^4)}({S}_{ee} + {S}_{\mu \mu}) \right],
\label{Zcharged}\\
&\Gamma(Z \to \ell_\alpha \ell_\beta) \simeq \frac{ m_Z^3}{3 \pi v^2 (16\pi^2)^2} \left( \frac{17 +t_w^2}{6} \right)^2 \left| \hat{R}_{\alpha \beta} \right|^2,
\label{widthZab}
\end{align}
and leptons flavor-violating decays are given by
\begin{align}
BR(\ell_\alpha \to \ell_\beta \gamma) & \simeq \frac{\alpha_{em} m_\alpha^5}{16  (16\pi^2)^2  \Gamma_\alpha} \left| S_{\alpha\beta}\right|^2,
\label{raddec}\\
BR(\ell_\alpha^- \to \ell_\beta^- \ell_\beta^+ \ell_\beta^-) 
&\simeq \frac{m_\alpha^5 \left(27-96s_w^2+128s_w^4 \right)}{36 \pi v^4 (16\pi^2)^3 \Gamma_\alpha} \left| \hat{R}_{\alpha \beta} \right|^2,
\label{3bodydec}
\end{align}
where $s_w=\sin\theta_w$, $t_w= \tan \theta_w$, $\theta_w$ is Weinberg angle, $v$ is the vacuum expectation value of the Higgs field. These relations were obtained due to seesaw contribution to low-energy four-fermion
operators or due to operators induced in the one-loop approximation in the renormalization group approach.  One-loop matching is performed at the scale of RH neutrinos, capturing important effects such as dipole transitions, see details in \cite{Coy:2018bxr}. 
\begin{table}[t]
    \centering
 \begin{tabular}{ |c|c|c|c|c| } 
 \hline
   & \multicolumn{2}{c|}{Present experiments } & \multicolumn{2}{c|}{Future experiments}\\
\hline
 quantity & observable  & upper limit\ & observable & upper limit \\
\hline
 $\hat S_{ee}+\hat S_{\mu \mu}$ &  $\Gamma(Z \to e^+ e^-) $ & $0.53\cdot 10^{-3}$ & - & -\\ 
\hline
 $\hat S_{\tau \tau}$ & $G^{\mu \tau}_{F}/G_{F}$ & $0.64\cdot 10^{-3}$ & - & - \\
\hline
$|\hat S_{e \mu}|$&  $BR(\mu \to e \gamma)$ & $6.8\cdot 10^{-6}$ & $BR(\mu \to e \gamma)$ &  $2.6\cdot 10^{-6}$ \\
\hline
 $|\hat S_{e \tau}|$& $BR(\tau \to e \gamma)$ & $4.5\cdot 10^{-3}$ & $BR(\tau \to e \gamma)$ & $1.8\cdot 10^{-3}$ \\
\hline
$|\hat S_{\mu \tau}|$& $BR(\tau \to \mu \gamma)$ &  $5.2\cdot 10^{-3}$ & $BR(\tau \to \mu \gamma)$ & $1.4\cdot 10^{-3}$ \\
\hline
 $|\hat R_{e \mu}|$ &  
    $  BR(\mu\, Au \to e\, Au)$
         & $
        9.7\cdot 10^{-6}
$ & $BR(\mu\, Ti \to e\, Ti)$ & $1.7\cdot 10^{-8}$ \\
\hline
 $|\hat R_{e \tau}|$ & $BR(\tau \to e e e)$ & 0.022 & $BR(\tau \to e e e)$ & $3.0\cdot 10^{-3}$\\
\hline
 $|\hat R_{\mu \tau}|$ & $BR(\tau \to \mu \mu \mu)$ & 0.019 & $BR(\tau \to \mu \mu \mu)$ & $4.2\cdot 10^{-3}$\\
\hline
\end{tabular}
    \captionsetup{justification=justified, singlelinecheck=false}
    \caption{ Upper bounds on the seesaw parameters $\hat S_{\alpha\beta}$ and $\hat R_{\alpha\beta}$ from present and expected in the foreseeable future experiments.}
    \label{tab:my_label1}
\end{table}

There are upper bounds on the parameters of the Lagrangian  \eqref{g1} resulting from measuring branching ratios for various processes with lepton flavour violation (e.g., $Z \to e\mu$, $\mu \to e\gamma$, etc), or measuring errors to observable reactions  (e.g., $Z \to 
 \ell^+ \ell^-$), see, e.g., \cite{Fernandez-Martinez:2016lgt,Coy:2018bxr,Zhang:2021jdf,Blennow:2023mqx}. 
The experimental constraints (upper bounds) for lepton flavor violating operator coefficients $\hat S_{\alpha\beta}=M_W^2 S_{\alpha\beta}$, $\hat R_{\alpha\beta}=M_W^2 R_{\alpha\beta}$, where $M_W$ is the $W$-boson mass, are presented in Tabl.\ref{tab:my_label1}, see details in \cite{Coy:2018bxr}. All the observables of interest depend solely on a single operator, either $\hat S_{ab}$ or $\hat R_{ab}$, and are therefore effectively “one-at-a-time bounds". The only exception is the correction to the $Z$-boson decay into $e^+e^-$, which is proportional to the sum $\hat S_{ee} + \hat S_{\mu\mu}$.

Neutrino modification of SM induces dipole and Yukawa-type operators in the SMEFT Lagrangian. 
These operators are encoded into $S_{\alpha\beta}$ and $R_{\alpha\beta}$ form because relevant observables are proportional to them. This parametrization automatically takes into account possible cancellations between operators of different types.
The contribution from the box diagrams $\propto F^4/(16\pi^2 M_I^2)$ to the penguin operators can be neglected for the region of mass and mixing angles that are considered in the paper. These diagrams become relevant close to the perturbativity regions as demonstrated in the recent paper \cite{Ruchayskiy:2022}.

Among the extensive literature discussing leptogenesis and LFV processes, only in some of them
 \cite{Drewes:2021nqr,Ruchayskiy:2022,Granelli:2022eru} LFV constraints were considered in the context of generating the required lepton asymmetry to produce the baryon asymmetry observed in the early Universe. It was shown that LFV constraints generally do not significantly restrict the parameter space for leptogenesis. However, it is interesting how small the relevant EFT operators can be while still enabling these theories to generate the observed baryon asymmetry of the Universe.

In this paper, we will not solve renormalization group equations and will not compute Feynman diagrams of processes. 
We will consider the existing limits on the observed parameters ($\hat S_{\alpha\beta}$ and $\hat R_{\alpha\beta}$) of the SM neutrino extension as a fact and 
consider a relation between these experimentally observable parameters. This will allow us to anticipate improvements in some experimental constraints.  We also want to see how far the upper bounds on the Lagrangian parameters are from the lower bounds resulting from the requirement for the generation of baryon asymmetry in the $\nu$MSM. To do this, we 
 express the baryon asymmetry of the early Universe in the $\nu$MSM via experimentally observed parameters and consider the lower bounds for them.

\section{Yukawa matrix parametrisation for 2 RH neutrinos}

In order to study the manifestation of sterile neutrinos at collider experiments, we can safely consider a case of two RH neutrinos modification of the SM. 
Elements of Yukawa matrix $F_{\alpha I}$ in the extended SM Lagrangian \eqref{g1} can be expressed via parameters of active neutrinos known from the data of active (left-handed, LH) neutrino oscillations. Most suitably, this can be done with the help of the Casas-Ibarra parameterization\footnote{There are also other approaches to parameterization, see \cite{Gorkavenko:2009vd}.} \cite{Casas:2001sr}:
\begin{align}\label{CasasIbarraDef}
F=\frac{i}{v}U_\nu\sqrt{m_\nu^{\rm diag}}\mathcal{R}\sqrt{M^{\rm diag}},
\end{align}
where  $(m_\nu^{\rm diag})_{ij}=\delta_{ij} m_i$ represents the diagonal matrix of 3 active neutrino masses,  $m_i$ are the masses of the active neutrinos. The matrix of active neutrino mixing in both flavor and mass bases $U_\nu$ can be written as:
\begin{align}
\label{PMNS}
U_\nu=V^{(23)}U_\delta V^{(13)}U_{-\delta}V^{(12)}{\rm diag}(e^{i \alpha_1/2},e^{i \alpha_2 /2},1)\,,
\end{align}
with $U_{\pm \delta}={\rm diag}(e^{\mp i \delta/2},1,e^{\pm i \delta /2})$ and
\begin{equation}
\label{VDef}
    V^{(12)}\!=\!
    \begin{pmatrix}
     c_{12} & s_{12} & 0\\
    -s_{12} & c_{12} & 0\\
     0      & 0      & 1\\
    \end{pmatrix},
\quad
 V^{(13)}\!=\!
    \begin{pmatrix}
     c_{13} & 0 & s_{13}\\
     0      & 1 & 0     \\
    -s_{13} & 0 & c_{13}\\
    \end{pmatrix},
\quad    
V^{(23)}\!=\!
    \begin{pmatrix}
    1  &  0      & 0     \\
    0  &  c_{23} & s_{13}\\
    0  & -s_{23} & c_{23}\\
    \end{pmatrix},    
\end{equation}
where $\delta$ is the Dirac phase, $\alpha_{1,2}$ are the Majorana phases, $\sin \theta_{ij}\equiv s_{ij}$, $\cos \theta_{ij}\equiv c_{ij}$, and the parameters $\theta_{ij}$ are the active neutrino mixing angles.

The complex orthogonal matrix $\mathcal{R}$ satisfies the condition $\mathcal{R}\mathcal{R}^T=1$. 
For the case of two RH neutrinos, there is only one complex angle $\omega$ for matrix $\mathcal{R}$ parameterization, and one has to distinguish between the cases of normal ordering (NO) and inverted ordering (IO) of active neutrino masses as follows:
\begin{align}
\label{RDef2}
\mathcal{R}^{\rm NO}=
\begin{pmatrix}
0 && 0\\
\cos \omega && \sin \omega \\
-\xi \sin \omega && \xi \cos \omega
\end{pmatrix}\,,\quad 
\mathcal{R}^{\rm IO}=
\begin{pmatrix}
\cos \omega && \sin \omega \\
-\xi \sin \omega && \xi \cos \omega \\
0 && 0
\end{pmatrix}
\,,
\end{align}
where $\xi=\pm 1$. As pointed out in \cite{Tastet:2021vwp}, changing the sign of $\xi$ can be undone by $\omega\rightarrow -\omega$ along with $N_2 \rightarrow -N_1$ \cite{Abada:2006ea}, so we set $\xi=+1$.
In the case of the normal neutrino hierarchy, we have 
\begin{equation}\label{normalIO}
    m_1=0,\quad m_2=\sqrt{\Delta m^2_{sol}}\approx 0.009\, {\rm eV},\quad m_3=\sqrt{\Delta m^2_{atm}}\approx 0.05\, {\rm eV}.
\end{equation}
In the case of the inverted neutrino hierarchy, we have 
\begin{equation}\label{invertedIO}
    m_3=0,\quad m_1\approx m_2 =\sqrt{\Delta m^2_{atm}}\approx 0.05\, {\rm eV}.
\end{equation}

Using the Casas-Ibarra parametrization \eqref{CasasIbarraDef},  one can easily obtain expressions for the elements of the Yukawa matrix $F_{\alpha I}$ for both normal and inverted ordering of active neutrino masses.

Normal ordering:
\begin{flalign}\label{Fe1Norm}
    &\quad F_{e1}=\frac{i \sqrt{M_1}}{v} \left(\sqrt{m_2} e^{\frac{i \alpha_2}{2}} \cos \theta_{13} \sin \theta_{12} \cos \omega -\sqrt{m_3}e^{-i \delta} \sin \theta_{13} \sin \omega \right),&&\\
\label{Fe2Norm}
   &\quad F_{e2}=\frac{i \sqrt{M_2}}{v} \left(\sqrt{m_2} e^{\frac{i \alpha_2}{2}} \cos \theta_{13} \sin \theta_{12} \sin \omega+ \sqrt{m_3} e^{-i \delta} \sin \theta_{13}\cos \omega\right),&&  \\  
\label{Fm1Norm}
    &\quad F_{\mu 1}=\frac{i \sqrt{M_1}}{v} \Bigl(-\sqrt{m_3} \cos \theta_{13}  \sin \theta_{23} \sin \omega +&& \nonumber\\
    &\qquad \qquad +\sqrt{m_2}e^{\frac{i \alpha_2}{2}}  \left(\cos \theta_{12}  \cos \theta_{23} -e^{i \delta} \sin \theta_{12} \sin \theta_{13} \sin \theta_{23}\right)\cos \omega\Bigr),&&\\
\label{Fm2Norm}
    &\quad F_{\mu 2}=\frac{i \sqrt{M_2}}{v} \Bigl(\sqrt{m_3}\cos \theta_{13} \sin \theta_{23} \cos \omega +&& \nonumber\\
    &\qquad \qquad +\sqrt{m_2}e^{\frac{i \alpha_2}{2}} \left(\cos \theta_{12} \cos \theta_{23} -e^{i \delta} \sin \theta_{12} \sin \theta_{13} \sin \theta_{23}\right)\sin \omega \Bigr),&&\\
\label{Ft1Norm}
    &\quad F_{\tau 1}=\frac{i \sqrt{M_1}}{v} \Bigl(-\sqrt{m_3} \cos \theta_{13} \cos \theta_{23} \sin \omega +&& \nonumber\\
    &\qquad \qquad +\sqrt{m_2} e^{\frac{i \alpha_2}{2}} \left(-\cos \theta_{12} \sin \theta_{23}-\cos \theta_{23} e^{i \delta} \sin \theta_{12} \sin \theta_{13}\right)\cos \omega \Bigr),&& \\
\label{Ft2Norm}
    &\quad F_{\tau 2}=\frac{i \sqrt{M_2}}{v} \Bigl(\sqrt{m_3} \cos \theta_{13} \cos \theta_{23}  \cos \omega +&& \nonumber\\
    &\qquad \qquad +\sqrt{m_2} e^{\frac{i \alpha_2}{2}} \left(-\cos \theta_{12} \sin \theta_{23}-\cos \theta_{23} e^{i \delta} \sin \theta_{12} \sin \theta_{13}\right)\sin \omega \Bigr).&&
\end{flalign}

Inverted ordering:
\begin{flalign}\label{Fe1Inv}
    &\quad F_{e1}=\frac{i \sqrt{M_1}}{v} \left(\sqrt{m_1} e^{\frac{i \alpha_1}{2}} \cos \theta_{13} \cos \theta_{12} \cos \omega -\sqrt{m_2}e^{\frac{i \alpha_2}{2}} \cos \theta_{13} \sin \theta_{12} \sin \omega \right),&&\\
\label{Fe2Inv}
   &\quad F_{e2}=\frac{i \sqrt{M_2}}{v} \left(\sqrt{m_1} e^{\frac{i \alpha_1}{2}} \cos \theta_{13} \cos \theta_{12} \sin \omega+ \sqrt{m_2} e^{\frac{i \alpha_2}{2}} \sin \theta_{12} \cos \theta_{13}\cos \omega\right),&&\\
    &\quad F_{\mu 1}=\frac{i \sqrt{M_1}}{v} \Bigl(\sqrt{m_1}e^{\frac{i \alpha_1}{2}} \left(-\cos \theta_{23}  \sin \theta_{12} -e^{i \delta} \cos \theta_{12} \sin \theta_{13} \sin \theta_{23}\right) \cos \omega -&& \nonumber\\
    &\qquad \qquad -\sqrt{m_2}e^{\frac{i \alpha_2}{2}}  \left(\cos \theta_{12}  \cos \theta_{23} -e^{\delta} \sin \theta_{12} \sin \theta_{13} \sin \theta_{23}\right)\sin \omega\Bigr),&&\label{Fm1Inv}\\
     &\quad F_{\mu 2}=\frac{i \sqrt{M_2}}{v} \Bigl(\sqrt{m_1}e^{\frac{i \alpha_1}{2}}\left(-\cos \theta_{23} \sin \theta_{12} -e^{i \delta} \cos \theta_{12} \sin \theta_{13} \sin \theta_{23}\right) \sin \omega +&& \nonumber\\
    &\qquad \qquad +\sqrt{m_2}e^{\frac{i \alpha_2}{2}} \left(\cos \theta_{12} \cos \theta_{23} -e^{i \delta} \sin \theta_{12} \sin \theta_{13} \sin \theta_{23}\right)\cos \omega \Bigr),&&\label{Fm2Inv}\\
    &\quad F_{\tau 1}=\frac{i \sqrt{M_1}}{v} \Bigl(\sqrt{m_1}e^{\frac{i \alpha_1}{2}} \left(\sin \theta_{12} \sin \theta_{23}-e^{i \delta}\cos \theta_{23}  \cos \theta_{12} \sin \theta_{13}\right) \cos \omega -&& \nonumber\\
    &\qquad \qquad -\sqrt{m_2} e^{\frac{i \alpha_2}{2}} \left(-\cos \theta_{12} \sin \theta_{23}-e^{i \delta} \cos \theta_{23} \sin \theta_{12} \sin \theta_{13}\right)\sin \omega \Bigr),&&\label{Ft1Inv}
 \end{flalign}  
\begin{flalign}
    &\quad F_{\tau 2}=\frac{i \sqrt{M_2}}{v} \Bigl(\sqrt{m_1}e^{\frac{i \alpha_1}{2}} \left(\sin \theta_{12} \sin \theta_{23}- e^{i \delta}\cos \theta_{23} \cos \theta_{12} \sin \theta_{13}\right) \sin \omega +&& \nonumber\\
    &\qquad \qquad +\sqrt{m_2} e^{\frac{i \alpha_2}{2}} \left(-\cos \theta_{12} \sin \theta_{23}-e^{i \delta} \cos \theta_{23} \sin \theta_{12} \sin \theta_{13}\right)\cos \omega \Bigr).&&\label{Ft2Inv}
\end{flalign}

\section{Relations between elements $S_{\alpha \beta}$ and  $R_{\alpha \beta}$}
Usage of the explicit relation for Yukawa matrix $F$ \eqref{Fe1Norm} -- \eqref{Ft2Inv} is very cumbersome for subsequent computations and analytical analysis. Thus, we introduce a new $3 \times 3$ complex  matrix 
\begin{equation}\label{Xmatrix}
  X=\frac{i}{v}U_\nu\sqrt{m_\nu^{\rm diag}}, 
\end{equation}
which enables us to write Yukawa matrix $F$ \eqref{CasasIbarraDef} in the form
\begin{equation}\label{FformX}
   F=X\mathcal{R}\sqrt{M^{\rm diag}}.
\end{equation}
Using this denotation and \eqref{RDef2}, we derived expressions for the observable matrix elements $S_{\alpha \beta}$ \eqref{Sab} for normal and inverted ordering of active neutrino masses in the case of extension of the SM by two RH neutrinos for arbitrary values of the model parameters. These relations are given in the Appendix \ref{Append_Sab_Rab}.

Let us now estimate the values of the complex parameter $w$. We can easily do it for the case of two heavy RH neutrinos with degenerated values of masses (as in $\nu$MSM model) $M \approx M_1 \approx M_2$.
In this case, there is a widely used parameter  \cite{Klaric:2021cpi,Ruchayskiy:2022}
\begin{equation}\label{UtotDef}
    U^2_{tot} =\sum_{\alpha,I} |\Theta_{\alpha I}|^2= \frac{v^2}{M^2}{\rm tr}(FF^\dagger) = \frac{\sum_i m_i}{M}\cosh2\mathfrak{Im}\omega,
\end{equation}
where $\Theta_{\alpha I}$ is the mixing angle between LH ($\nu_\alpha$) and  RH neutrinos ($N_I$), $M$ is the mass of the RH neutrinos.
Given that  mixing angles are finite and masses $m_\alpha$ of active neutrinos are extremely small compared to masses of right-handed neutrinos $M_I$, 
we will consider an interesting region for the current experimental search of HNL above the sea-saw line when
\begin{equation}\label{BigOmegaAssum}
    \cosh2\mathfrak{Im}\omega \approx \sinh2\mathfrak{Im}\omega \approx \frac{\exp{2\mathfrak{Im}\omega}}{2}{\gg 1}.
\end{equation}
This assumption will also be valid for different heavy neutrino masses.

The assumption \eqref{BigOmegaAssum} allows us to derive fairly simple relations for experimentally observable elements $S_{\alpha \beta}$ and  $R_{\alpha \beta}$.

 Normal ordering of active neutrino masses:
 \begin{equation}\label{SeeNorm}
  S_{ee}=\frac{e^{2 \mathfrak{Im}\omega}}{4}\frac{ M_1+M_2}{ M_1 M_2} (X_{12}-i X_{13}) \left(X_{12}^*+i X_{13}^*\right) ,
 \end{equation}
 \begin{equation}\label{SmumuNorm}
  S_{\mu \mu}=\frac{e^{2 \mathfrak{Im}\omega}}{4}\frac{ M_1+M_2}{ M_1 M_2}(X_{22}-i X_{23}) \left(X_{22}^*+i X_{23}^*\right) ,
 \end{equation}
 \begin{equation}\label{StautauNorm}
  S_{\tau \tau}=\frac{e^{2 \mathfrak{Im}\omega}}{4}\frac{ M_1+M_2}{ M_1 M_2}(X_{32}-i X_{33}) \left(X_{32}^*+i X_{33}^*\right),
 \end{equation}
 \begin{equation}\label{SemuNorm}
  S_{e \mu}=\frac{e^{2 \mathfrak{Im}\omega}}{4}\frac{ M_1+M_2}{ M_1 M_2}(X_{12}-i X_{13}) \left(X_{22}^*+i X_{23}^*\right),
 \end{equation}
 \begin{equation}\label{SetauNorm}
  S_{e \tau}=\frac{e^{2 \mathfrak{Im}\omega}}{4}\frac{ M_1+M_2}{ M_1 M_2}(X_{12}-i X_{13}) \left(X_{32}^*+i X_{33}^*\right),
 \end{equation}
 \begin{equation}\label{SmutauNorm}
  S_{\mu \tau}=\frac{e^{2 \mathfrak{Im}\omega}}{4}\frac{ M_1+M_2}{ M_1 M_2}(X_{22}-i X_{23}) \left(X_{32}^*+i X_{33}^*\right).
 \end{equation}

Inverted ordering of active neutrino
masses:
\begin{equation}\label{SeeInv}
  S_{ee}=\frac{e^{2 \mathfrak{Im}\omega}}{4}\frac{ M_1+M_2}{ M_1 M_2} (X_{11}-i X_{12}) \left(X_{11}^*+i X_{12}^*\right),
 \end{equation}
 \begin{equation}\label{SmumuInv}
  S_{\mu \mu}=\frac{e^{2 \mathfrak{Im}\omega}}{4}\frac{ M_1+M_2}{ M_1 M_2}(X_{21}-i X_{22}) \left(X_{21}^*+i X_{22}^*\right),
 \end{equation}
 \begin{equation}\label{StautauInv}
  S_{\tau \tau}=\frac{e^{2 \mathfrak{Im}\omega}}{4}\frac{ M_1+M_2}{ M_1 M_2}(X_{31}-i X_{32}) \left(X_{31}^*+i X_{32}^*\right),
 \end{equation}
 \begin{equation}\label{SemuInv}
  S_{e \mu}=\frac{e^{2 \mathfrak{Im}\omega}}{4}\frac{ M_1+M_2}{ M_1 M_2}(X_{11}-i X_{12}) \left(X_{21}^*+i X_{22}^*\right),
 \end{equation}
 \begin{equation}\label{SetauInv}
  S_{e \tau}=\frac{e^{2 \mathfrak{Im}\omega}}{4}\frac{ M_1+M_2}{ M_1 M_2}(X_{11}-i X_{12}) \left(X_{31}^*+i X_{32}^*\right),
 \end{equation}
 \begin{equation}\label{SmutauInv}
  S_{\mu \tau}=\frac{e^{2 \mathfrak{Im}\omega}}{4}\frac{ M_1+M_2}{ M_1 M_2}(X_{21}-i X_{22}) \left(X_{31}^*+i X_{32}^*\right).
 \end{equation}
As one can effortlessly see, diagonal elements $S_{\alpha\alpha}$ are real and positive. 
We would like to point out the fact that under physically reasonable assumption \eqref{BigOmegaAssum} relations \eqref{SeeNorm} -- \eqref{SmutauInv} are independent of $\mathfrak{Re}\omega$, so the number of parameters is decreased by 1. This can be comprehended by examining the general expressions provided in the Appendix \ref{Append_Sab_Rab}.

 Using explicit form for $S_{\alpha\beta}$ \eqref{SeeNorm} -- \eqref{SmutauInv}, one can analytically obtain relations between observable matrix elements  $S_{\alpha \beta}$ \eqref{Sab}, $R_{\alpha \beta}$ \eqref{Rab}, that are valid  in the case of different mass values  of two RH neutrinos, massive active neutrinos, and under assumption \eqref{BigOmegaAssum}
\begin{equation}\label{SandR}
    S_{\alpha \beta}\Bigl(M_1\ln\frac{M_2}{M_W}+M_2\ln\frac{M_1}{M_W}\Bigr)=R_{\alpha \beta}(M_1+M_2).
\end{equation}
Using explicit form for $S_{\alpha\beta}$ \eqref{SeeNorm} -- \eqref{SmutauInv} and relation \eqref{SandR}, we can establish strong general constraints on $S_{\alpha \beta}$, $R_{\alpha \beta}$
\begin{equation}\label{SRrelations}
    |S_{\alpha \beta}|^2= S_{\alpha \alpha} S_{\beta \beta} , \quad
    |R_{\alpha \beta}|^2 =  R_{\alpha \alpha} R_{\beta \beta},
\end{equation}
that are valid in the case of different mass values of two RH neutrinos and massive active neutrinos. These relations do not depend on the mass scale of the RH neutrino. The only thing that is required is that the masses of the active neutrinos be extremely small compared to the masses of the right-handed neutrinos and the condition \eqref{BigOmegaAssum} be satisfied.

We derive analytically that Schwarz inequalities $|S_{\alpha \beta}|^2 \leq S_{\alpha \alpha} S_{\beta \beta} $ and $|R_{\alpha \beta}|^2 \leq R_{\alpha \alpha} R_{\beta \beta} $ saturate only when $e^{\mathfrak{Im}\omega}\gg 1$. More precisely, in the case of finite large values of $e^{\mathfrak{Im}\omega}$ we get
\begin{equation}
    \frac{|S_{\alpha \beta}|^2 - S_{\alpha \alpha} S_{\beta \beta}}{S_{\alpha \alpha} S_{\beta \beta}}\sim e^{-4\mathfrak{Im}\omega},\quad  
    \frac{|R_{\alpha \beta}|^2 - R_{\alpha \alpha} R_{\beta \beta}}{R_{\alpha \alpha} R_{\beta \beta}}\sim e^{-4\mathfrak{Im}\omega}, \quad e^{\mathfrak{Im}\omega} \gg 1.
\end{equation}
Since to prove the above statement we used the $X$ matrix \eqref{Xmatrix} proportional to the mass of active neutrinos our statement is valid only for different (possibly close) values of $M_1$ and $M_2$ masses. The independence of the obtained result from the difference in mass of sterile neutrinos is very important. This is because the commonly used case of almost degenerate values of the masses of RH neutrinos is still only a theoretical assumption.

Restrictions \eqref{SRrelations} are important because previously similar restrictions in the form of saturated Schwarz inequality have been obtained only for the partial case of massless active neutrinos  or 
degenerate masses of heavy sterile neutrinos \cite{Coy:2018bxr,Blennow:2023mqx}.

Assuming that the restrictions \eqref{SRrelations} are satisfied with sufficient accuracy for the condition \eqref{BigOmegaAssum} above the see-saw line  we find constraints on the tau-flavours from much stronger constraints on muon and electron flavors. It turns out that our predictions provide better constraints than the experimental data for $\hat S_{e\tau}$ and $\hat S_{\mu\tau}$ elements  by an order of magnitude, see Tabl.\ref{tab:my_label2}.

\begin{table}[t]
    \centering
 \begin{tabular}{ |c|c|c|c| } 
\hline
quantity & experimental & improved \\
\hline
 $|\hat S_{e \tau}|$& $4.5\cdot 10^{-3} (1.8\cdot 10^{-3})$ & $0.58\cdot 10^{-3}$ \\
 \hline
$|\hat S_{\mu \tau}|$& $5.2\cdot 10^{-3} (1.4\cdot 10^{-3})$ & $0.58\cdot 10^{-3}$ \\
\hline
\end{tabular}
    \captionsetup{justification=justified, singlelinecheck=false}
    \caption{Experimental and newly calculated upper bounds on the seesaw parameters $\hat S_{\alpha\beta}$ and $\hat R_{\alpha\beta}$. Values in brackets correspond to constraints from expected in the foreseeable future experiments.}
    \label{tab:my_label2}
\end{table}

Using definition  $X=\frac{i}{v}U_\nu\sqrt{m_\nu^{\rm diag}}$  and \eqref{SeeNorm} -- \eqref{SmutauInv}, we can obtain explicit expressions for the observable diagonal elements $S_{\alpha\alpha}$ for normal and inverted ordering of active neutrino masses via parameters of active neutrino oscillations, $\mathfrak{Im}\omega$, the masses of two heavy RH neutrinos and under the condition \eqref{BigOmegaAssum}.

 Normal ordering:
\begin{multline}\label{SeeCasasIbarraNorm}
 S_{ee}= \frac{e^{2 \mathfrak{Im}\omega}}{4v^2}\frac{M_1+M_2}{ M_1 M_2 }\times \\ \left( \vphantom{\frac12} m_2 \sin ^2\theta_{12} \cos ^2\theta_{13}+m_3 \sin ^2\theta_{13}
 -\sqrt{m_2 m_3} \sin\theta_{12} \sin2 \theta_{13} \sin \frac{\alpha_2+2\delta}{2}\right),
\end{multline}
\vspace{-1em}
\begin{multline}\label{SmumuCasasIbarraNorm}
   S_{\mu \mu} =  \frac{e^{2 \mathfrak{Im}\omega}}{4v^2}\frac{M_1+M_2}{ M_1 M_2 } \left(\frac{1}{4} \sin ^2\theta_{23} \left[ \vphantom{\frac12} 4 \sqrt{m_2 m_3} \sin \theta_{12} \sin (2 \theta_{13}) \sin \frac{\alpha_2+2\delta}{2}-\right.\right.\\ \left.
   (m_2-2 m_3) \cos 2 \theta_{13}-2 m_2 \cos 2 \theta_{12} \sin ^2\theta_{13}+m_2+2 m_3 \vphantom{\frac12}\right]-\\ 
   2 \cos \theta_{12} \sin \theta_{23} \cos \theta_{23} \left[ \vphantom{\frac12}\sin \frac{\alpha_2}{2} \sqrt{m_2 m_3} \cos \theta_{13} \left.
   +m_2 \cos \delta \sin \theta_{12} \sin \theta_{13} \vphantom{\frac12}\right]+m_2 \cos ^2\theta_{12} \cos ^2\theta_{23}\right),
\end{multline}
\vspace{-1em}
\begin{multline}\label{StautauCasasIbarraNorm}
   S_{\tau \tau} = \frac{e^{2 \mathfrak{Im}\omega}}{4v^2}\frac{M_1+M_2}{ M_1 M_2 } \left( \vphantom{\frac12} \sin \frac{\alpha_2}{2} \sqrt{m_2 m_3} \cos  \theta_{12} \cos\theta_{13} \sin 2 \theta_{23}+\right.\\ \left.
   \sin \theta_{12} \cos ^2\theta_{23} \left[\sqrt{m_2 m_3} \sin 2 \theta_{13} \sin \frac{\alpha_2 +2\delta}{2}
   +m_2 \sin \theta_{12} \sin ^2\theta_{13}\right]+\right.\\ \left.
   2 m_2 \cos \delta \sin \theta_{12} \cos \theta_{12} \sin \theta_{13} \sin \theta_{23} \cos \theta_{23}  +m_2 \cos ^2\theta_{12} \sin ^2\theta_{23}+m_3 \cos ^2\theta_{13} \cos ^2\theta_{23} \vphantom{\frac12} \right).
\end{multline}

 Inverted ordering:
 \begin{multline}\label{SeeCasasIbarraInv}
    S_{ee}= \frac{e^{2 \mathfrak{Im}\omega}}{4v^2}\frac{M_1+M_2}{ M_1 M_2 }\times\\ \cos ^2\theta_{13} \left( \vphantom{\frac12} m_1 \cos ^2\theta_{12}+m_2 \sin ^2\theta_{12}
    -\sqrt{m_1 m_2} \sin 2 \theta_{12} \sin \frac{\alpha_1-\alpha_2}{2} \vphantom{\frac12} \right),
 \end{multline}
 \vspace{-1em}
\begin{multline}\label{SmumuCasasIbarraInv}
     S_{\mu \mu} = \frac{e^{2 \mathfrak{Im}\omega}}{8v^2}\frac{M_1+M_2}{ M_1 M_2 }
     \left (\vphantom{\frac12}\sin \theta_{13} \left( \sin 2 \theta_{23} \left[2 \sqrt{m_1 m_2} \left \{\sin \delta \cos \frac{\alpha_1-\alpha_2}{2}+\right.\right.\right.\right.\\ \left. \left.
     \cos \delta \cos 2 \theta_{12} \sin \frac{\alpha_1-\alpha_2}{2} \right \}+m_1 \cos \delta \sin 2 \theta_{12}\right]+
\\
     \left. \sin \theta_{13} \sin ^2\theta_{23} \left(-2 \sqrt{m_1 m_2} \sin 2 \theta_{12} \sin \frac{\alpha_1-\alpha_2}{2}
     +(m_1-m_2) \cos 2 \theta_{12}+m_1+m_2 \vphantom{\frac12}\right) \right)+\\
     \cos ^2\theta_{23} \left[ 2 \sqrt{m_1 m_2} \sin 2 \theta_{12} \sin \frac{\alpha_1-\alpha_2}{2}    +(m_2-m_1) \cos 2 \theta_{12}+m_1+m_2 \vphantom{\frac12} \right]-\\  
     \left.
     -4 m_2 \cos \delta \sin \theta_{12} \cos \theta_{12} \sin \theta_{13} \sin \theta_{23} \cos \theta_{23}\vphantom{\frac12}\right),
 \end{multline}
 \vspace{-1em}
\begin{multline}\label{StautauCasasIbarraInv}
      S_{\tau \tau} =\frac{e^{2 \mathfrak{Im}\omega}}{8v^2}\frac{M_1+M_2}{ M_1 M_2 }
      \left(-4 \sin \theta_{13} \sin \theta_{23} \cos \theta_{23} \left[ \sin \delta \sqrt{m_1 m_2} \cos \frac{\alpha_1-\alpha_2}{2}+\right.\right.\\ \left. \left.
      \cos \delta (m_1-m_2) \sin \theta_{12} \cos \theta_{12}\vphantom{\frac12}\right]-\right.\\ \left.
      2 \cos \delta \sqrt{m_1 m_2} \cos 2 \theta_{12} \sin \theta_{13} \sin 2 \theta_{23} \sin \frac{\alpha_1-\alpha_2}{2}+\right.\\ \left.
      \sin ^2\theta_{13} \cos ^2\theta_{23} \left[-2 \sqrt{m_1 m_2} \sin 2 \theta_{12} \sin \frac{\alpha_1-\alpha_2}{2}+(m_1-m_2) \cos 2 \theta_{12}+m_1
      +m_2 \right]+\right.\\ \left.
      \sin ^2\theta_{23} \left[2 \sqrt{m_1 m_2} \sin 2 \theta_{12} \sin \frac{\alpha_1-\alpha_2}{2}+(m_2-m_1) \cos 2 \theta_{12}+m_1+m_2\right]\right).
 \end{multline}
 Explicit expressions for $R_{\alpha \alpha}$ can be easily obtained with the help of relation \eqref{SandR}. Explicit expressions for observables $|S_{\alpha \beta}|$ and $|R_{\alpha \beta}|$ can be  obtained with the help of   \eqref{SRrelations}.

\section{Numerical results and constraints}

Relations \eqref{SeeCasasIbarraNorm} -- \eqref{StautauCasasIbarraInv} contain factor $e^{2 \mathfrak{Im}\omega} (M_1+M_2)/( M_1 M_2)$. 
We can get rid of this factor with the help of relations 
\eqref{UtotDef}, \eqref{BigOmegaAssum} in the approximation $M_1\approx M_2 \approx M$
\begin{equation}\label{M1eqM2Approx}
 \frac{e^{2 \mathfrak{Im}\omega}}{4}\frac{ M_1+M_2 }{ M_1 M_2}\to \frac{U_{tot}^2}{\sum_i m_i}.
\end{equation}
In this approximation, relation \eqref{SandR} transforms into
\begin{equation}\label{RabSabApprox}
    R_{\alpha \beta}=S_{\alpha \beta} \ln \frac{M}{M_W}.
\end{equation}
We obtained numerical expressions for dimensionless quantities $\hat S_{\alpha \alpha}=M_W^2 S_{\alpha\beta}$, and $\hat R_{\alpha \alpha}=M_W^2 R_{\alpha\beta}$ by taking the best-fit values for active neutrino's parameters \cite{ParticleDataGroup:2022pth}.

Thus, for the case of normal ordering of active neutrino
masses, we get
\begin{equation}\label{SeeNumNorm}
   \hat S_{ee}=U_{tot}^2 \left(6.6\cdot 10^{-3} -6.1\cdot 10^{-3} \sin \frac{\alpha_2+8.5}{2}\right),
\end{equation}
\begin{equation}\label{SmumuNumNorm}
  \hat S_{\mu \mu}=U_{tot}^2 \left(5.4\cdot 10^{-2}-3.1\cdot 10^{-2} \sin  \frac{\alpha_2}{2}+3.3\cdot 10^{-3} \sin \frac{\alpha_2+8.5}{2} \right),
\end{equation}
\begin{equation}\label{StautauNumNorm}
    \hat S_{\tau \tau}=U_{tot}^2 \left(4.6\cdot 10^{-2} + 3.1\cdot 10^{-2} \sin \frac{\alpha_2}{2}+ 2.8\cdot 10^{-3} \sin \frac{\alpha_2+8.5}{2}\right).
\end{equation}

For the case of inverted ordering of active neutrino
masses, we get
\begin{equation}\label{SeeNumInv}
   \hat S_{ee}=U_{tot}^2 \left(5.2\cdot 10^{-2} -4.8\cdot 10^{-2} \sin \frac{\alpha_1-\alpha_2}{2}\right),
\end{equation}
\begin{equation}\label{SmumuNumInv}
  \hat S_{\mu \mu}=U_{tot}^2 \left(2.5\cdot 10^{-2}-7.1 \cdot 10^{-3} \cos \frac{\alpha_1-\alpha_2}{2}+2.1\cdot 10^{-2}  \sin \frac{\alpha_1-\alpha_2}{2}\right),
\end{equation}
\begin{equation}\label{StautauNumInv}
    \hat S_{\tau \tau}=U_{tot}^2 \left(2.9\cdot 10^{-2} + 7.1\cdot 10^{-3}  \cos \frac{\alpha_1-\alpha_2}{2}+ 2.7\cdot 10^{-2}  \sin \frac{\alpha_1-\alpha_2}{2} \right).
\end{equation}

Using Tabl.\ref{tab:my_label1}, we can derive upper bounds for quantities $U_{tot}^2$ and $U_{tot}^2\ln\frac{M}{M_W}$ from upper bounds on $\hat S_{\alpha \beta}$ and $\hat R_{\alpha \beta}$, respectively. It can be verified that the strongest bounds originate from $\hat S_{e \mu}$ and $\hat R_{e \mu}$. We thus obtain the following constraints for quantities $U_{tot}^2$ and $U_{tot}^2\ln\frac{M}{M_W}$ for the present (and expected in the foreseeable future) experimental upper bounds on $\hat S_{e \mu}=(\hat S_{ee} \hat S_{\mu\mu})^{1/2}$ and $\hat R_{e \mu}=\hat S_{e\mu} \ln \frac{M}{M_W}$.

For the case of normal ordering of active neutrino
masses,
\begin{multline}\label{UtotSNumNorm}
 \hat S_{e\mu} =   U_{tot}^2\left(5.4\cdot 10^{-2}-3.1\cdot 10^{-2} \sin \frac{\alpha_2}{2}+3.3\cdot 10^{-3} \sin \frac{\alpha_2+8.5}{2} \right)^{1/2}\\
   \times \left(6.6\cdot 10^{-3} -6.1\cdot 10^{-3} \sin \frac{\alpha_2+8.5}{2}\right)^{1/2}\leq 6.8\cdot 10^{-6} (2.6\cdot 10^{-6}),
\end{multline}
\begin{multline}\label{UtotRNumNorm}
   \hat R_{e\mu} =   U_{tot}^2 \ln\frac{M}{M_W} \left(5.4\cdot 10^{-2}-3.1\cdot 10^{-2} \sin \frac{\alpha_2}{2}+3.3\cdot 10^{-3} \sin \frac{\alpha_2+8.5}{2} \right)^{1/2}\\
   \times \left(6.6\cdot 10^{-3} -6.1\cdot 10^{-3} \sin \frac{\alpha_2+8.5}{2}\right)^{1/2}\leq 9.7\cdot 10^{-6} (1.7\cdot 10^{-8}).
\end{multline}

For the case of inverted ordering of active neutrino
masses,
\begin{multline}\label{UtotSNumInv}
 \hat S_{e\mu} =    U_{tot}^2 \left(2.5\cdot 10^{-2}-7.1 \cdot 10^{-3} \cos \frac{\alpha_1-\alpha_2}{2}+2.1\cdot 10^{-2}  \sin \frac{\alpha_1-\alpha_2}{2} \right)^{1/2}\\
   \times \left(5.2\cdot 10^{-2} -4.8\cdot 10^{-2} \sin \frac{\alpha_1-\alpha_2}{2}\right)^{1/2}\leq 6.8\cdot 10^{-6} (2.6\cdot 10^{-6}),
\end{multline}
\vspace{-1em}
\begin{multline}\label{UtotRNumInv}
\hat R_{e\mu} =      U_{tot}^2 \ln\frac{M}{M_W} \left(2.5\cdot 10^{-2}-7.1 \cdot 10^{-3} \cos \frac{\alpha_1-\alpha_2}{2}+2.1\cdot 10^{-2}  \sin \frac{\alpha_1-\alpha_2}{2} \right)^{1/2}\\
   \times \left(5.2\cdot 10^{-2} -4.8\cdot 10^{-2} \sin \frac{\alpha_1-\alpha_2}{2}\right)^{1/2}\leq 9.7\cdot 10^{-6} (1.7\cdot 10^{-8}).
\end{multline}

Minimizing the derived expressions above with the aid of Majorana phases $\alpha_1$ and $\alpha_2$ over the interval $[0,2\pi]$,
 we can obtain upper bounds for quantities $U_{tot}^2$ and $U_{tot}^2\ln\frac{M}{M_W}$.

 \begin{figure}[t]
    \centering
    \hspace{-1cm}\includegraphics[width = 1.05\linewidth]{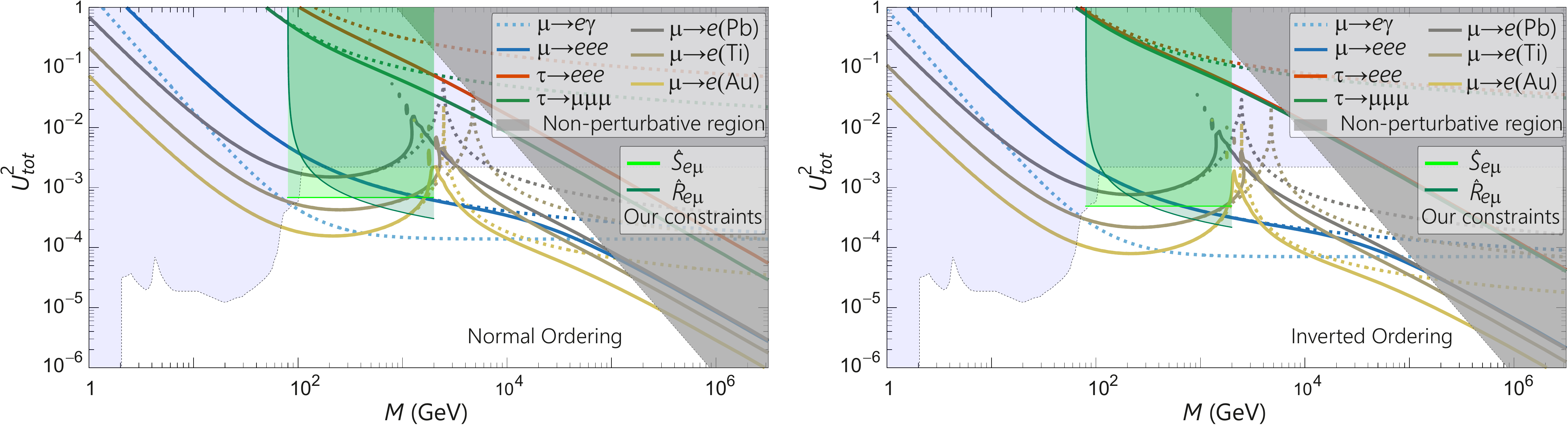}
    \captionsetup{justification=justified, singlelinecheck=false}    
    \caption{Imposing our restrictions on previously obtained restrictions from present experimental data, see details in \cite{Ruchayskiy:2022}, on
  HNLs parameters from charged lepton flavor violation processes. The light solid horizontal green line was computed from constraint on $\hat S_{e \mu}$. The dark solid green line (almost vertical in the vicinity of $M=M_W$)  - from our constraint on $\hat R_{e \mu}$, see \eqref{UtotNumNorm}, \eqref{UtotNumInv}. The painted regions of parameters are forbidden.
}
    \label{fig:coulomb_states}
\end{figure}

For the case of normal ordering of active neutrino
masses, we get
\begin{equation}\label{UtotNumNorm}
    U_{tot}^2 \leq 6.85\cdot 10^{-4} (2.62\cdot 10^{-4}), \quad
    U_{tot}^2 \ln\frac{M}{M_W} \leq 9.78\cdot 10^{-4} (1.71\cdot 10^{-6}).
\end{equation}

For the case of inverted ordering of active neutrino
masses, we get 
\begin{equation}\label{UtotNumInv}
    U_{tot}^2 \leq 4.91\cdot 10^{-4} (1.88\cdot 10^{-4}), \quad
    U_{tot}^2 \ln\frac{M}{M_W} \leq 7.00\cdot 10^{-4} (1.23\cdot 10^{-6}).
\end{equation}

\begin{figure}[t]
    \centering
    \hspace{-1cm}\includegraphics[width = 1.05\linewidth]{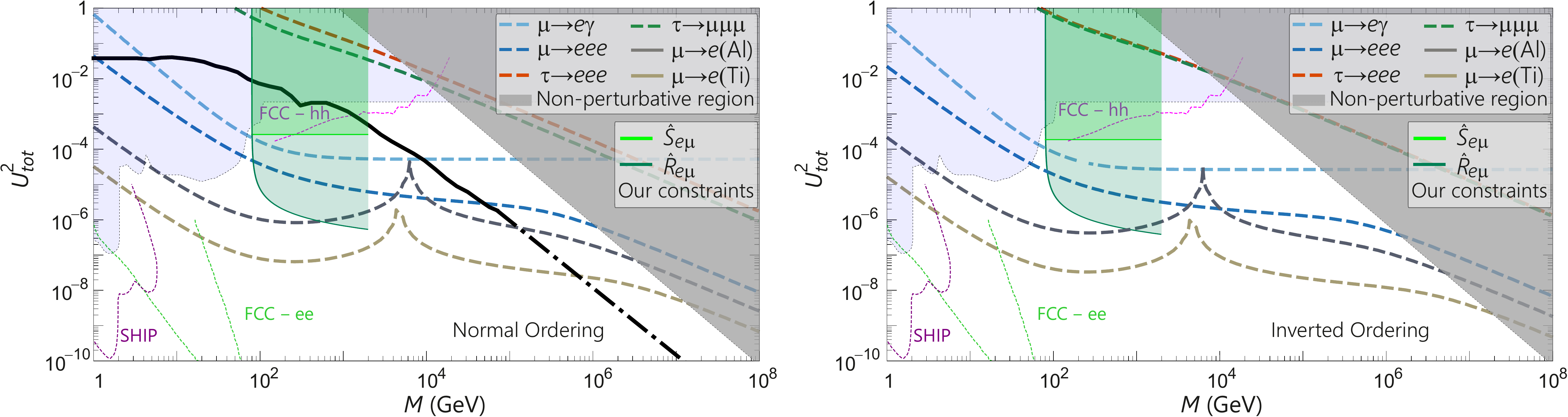}
    \captionsetup{justification=justified, singlelinecheck=false}
    \caption{Imposing our restrictions on previously obtained restrictions from expected in the foreseeable future experiments, see details in \cite{Ruchayskiy:2022}, on
  HNLs parameters from charged lepton flavor violation processes. The light solid horizontal green line was computed from our constraint on $\hat S_{e \mu}$. The dark solid green line (almost vertical in the vicinity of $M=M_W$)  - from the constraint on $\hat R_{e \mu}$, see \eqref{UtotNumNorm}, \eqref{UtotNumInv}. The painted regions of parameters are forbidden. The left plot's black line is the parameter space's upper boundary where successful leptogenesis with 3 RH neutrinos is
possible \cite{Drewes:2021nqr}.}\label{fig:coulomb_states2}
\end{figure}

Using these results, we have built graphs that illustrate the forbidden area for $M$ and $U_{tot}^2$ and compared them to the results obtained in \cite{Ruchayskiy:2022}, see Fig.\ref{fig:coulomb_states} and Fig.\ref{fig:coulomb_states2}. Starting relations \eqref{Zcharged} -- \eqref{3bodydec}  were obtained using renormalization group equations in one-loop approximation that gives quite a big error for the sterile neutrino masses larger than 10 TeV. So we considered only the region of the RH neutrino masses up to several TeV. As one can see, the simple constraints \eqref{UtotNumNorm}, \eqref{UtotNumInv} agree quite well
with the results of more complicated computations in the model with two RH neutrinos. 

\section{Baryon asymmetry}

In  \cite{Asaka:2005}, the generation of baryon asymmetry in the $\nu MSM$ 
was theoretically investigated. Baryon asymmetry was expressed through the total asymmetry
in the sterile neutrino sector
\begin{equation}\label{nB}
    \frac{n_B}{s}=7\cdot 10^{-4}  {\rm Tr} \Delta_N|_{T_W}=(8.8 - 9.8)\cdot 10^{-11},
\end{equation}
where
\begin{equation}\label{g11}
    {\rm Tr} \Delta_N|_{T_W} = \frac{\pi^{\frac32}\sin^3\phi}{384\cdot3^{\frac13}\Gamma(\frac56)}
    \frac{m^{\frac12}_{sol} m^{\frac32}_{atm} M_1^{\frac12} M_2^{\frac32} M_0^{\frac73}}{v^4 T_W (\Delta M^2_{21})^{\frac23}} \sum_{\alpha, I} a^\alpha |F_{\alpha I}|^2.
\end{equation}
Here, $T_W \approx 100\ {\rm GeV}$ is the electroweak phase transition temperature, $\sin\phi \approx 0.02$, $M_0 \simeq 7\cdot10^{17}\ {\rm GeV}$,  $\alpha$ -- lepton flavors, $I=1,2$ are indices of heavy neutrinos that contribute to baryon asymmetry, $M_{1,2}$ -- masses of these two heavy sterile neutrinos with almost degenerate values\footnote{In the original paper \cite{Asaka:2005}, slightly different notations were used, namely: the lightest sterile neutrino (dark matter candidate particle) there was $M_1$; two heavy neutrinos $M_2$ and $M_3$ were expressed in terms $f_2$ and $f_3$ correspondingly.}, $\Delta M^2_{21}=M_2^2-M_1^2$,
\begin{equation}\label{g9}
  a^\alpha =  4\dfrac{ {\rm Im}[F_{\alpha 3}(F^\dagger F)_{32} F^\dagger_{2\alpha}]}{f_1f_2(f_2^2-f_1^2)},\quad f_1^2 = \frac{m_{sol}M_1}{v^2}, \quad f_2^2 = \frac{m_{atm}M_2}{v^2}.
\end{equation}

Let us rewrite the parameter $ {\rm Im}[F_{\alpha 3}(F^\dagger F)_{32} F^\dagger_{2\alpha}]=\delta^\alpha$ in the form
\begin{equation}\label{delta_a_Yukawa}
    \begin{aligned}
        & \delta^e = {\rm Im}[F_{e2}F_{e1}^\ast (F_{\mu2}^\ast F_{\mu1}  +  F_{\tau2}^\ast F_{\tau1})],\\
        & \delta^\mu = {\rm Im}[F_{\mu2}F_{\mu1}^\ast (F_{e2}^\ast F_{e1}  +  F_{\tau2}^\ast F_{\tau1})],\\
        &\delta^\tau = {\rm Im}[F_{\tau2}F_{\tau1}^\ast (F_{e2}^\ast F_{e1}  +  F_{\mu2}^\ast F_{\mu1})],
    \end{aligned} 
    \Rightarrow 
    \delta^\alpha = \sum_{\beta\neq\alpha} {\rm Im} [(F_{\alpha1} F_{\beta1}^\ast)^\ast (F_{\alpha2} F_{\beta2}^\ast)].
\end{equation}
From the system of equations \eqref{Sab}, \eqref{Rab} one can derive
\begin{align}
\label{Fa1Fb1*}
    & F_{\alpha1} F_{\beta1}^\ast = \frac{M_1^2}{\ln\dfrac{M_2}{M_1}}\left(S_{\alpha\beta} \ln\dfrac{M_2}{M_W} - R_{\alpha\beta}\right),\\
\label{Fa2Fb2*}
    & F_{\alpha2} F_{\beta2}^\ast = \frac{M_2^2}{\ln\dfrac{M_2}{M_1}}\left(R_{\alpha\beta} - S_{\alpha\beta} \ln\dfrac{M_1}{M_W} \right).
\end{align}

It allows us to obtain the following expression in terms of the experimentally observed elements of the matrices $S$ \eqref{Sab} and $R$  \eqref{Rab}
\begin{multline}\label{Sum_deltaa_FF*}
    \sum_{\alpha, I} \delta^\alpha |F_{\alpha I}|^2  =
    \frac{M_1^2 M_2^2}{\ln^2\dfrac{M_2}{M_1}} \times \\
    \sum_\alpha  \left \{\sum_{\beta\neq\alpha} {\rm Im}\left[S_{\alpha\beta}^\ast R_{\alpha\beta}\right]
    \left[M_1^2 \left(S_{\alpha\alpha}\ln\frac{M_2}{M_W} - R_{\alpha\alpha}\right) + M_2^2 \left(R_{\alpha\alpha} - S_{\alpha\alpha}\ln\frac{M_1}{M_W}\right) \right] \right\}.
\end{multline}
Based on definitions \eqref{Sab}, and \eqref{Rab} one can see that diagonal elements $S_{\alpha \alpha}$, $R_{\alpha \alpha}$ are real. Using expression \eqref{SandR} it is easy to see that the combination $S_{\alpha \beta}^* R_{\alpha \beta}$ is also real. Hence, factor $\sum_{\alpha, I} \delta^\alpha |F_{\alpha I}|^2$, as defined by \eqref{Sum_deltaa_FF*}, is zero,  consequently leading to a vanishing baryon asymmetry of the Universe.

This obviously incorrect outcome stems from using approximation \eqref{BigOmegaAssum}.  Taking into account correct expressions for $S_{\alpha \beta}$, $R_{\alpha \beta}$, see Appendix \ref{Append_Sab_Rab},
 we can  rewrite the terms of expression \eqref{Sum_deltaa_FF*} as
\begin{equation}\label{Deltaa_FF_term2}
    \left[M_1^2 \left(S_{\alpha\alpha}\ln\frac{M_2}{M_W} - R_{\alpha\alpha}\right) + M_2^2 \left(R_{\alpha\alpha} - S_{\alpha\alpha}\ln\frac{M_1}{M_W}\right) \right]= M_1 M_2 S_{\alpha\alpha} \ln\frac{M_2}{M_1},
\end{equation}
In the case of normal ordering of active neutrinos, one can get
\begin{multline}\label{ImSem*Rem}
    {\rm Im}\left[S_{e\mu}^\ast R_{e\mu}\right]=\frac{\ln\frac{M_2}{M_1}}{M_1+M_2}{\rm Im}\left[S_{e \mu}^\ast \left \{(-X_{12}X_{22}^\ast+X_{13}X_{23}^\ast)\cos{2\rm \mathfrak{Re} \omega}+\right. \right.\\ \left. \left.
    +(X_{13}X_{22}^\ast+X_{12}X_{23}^\ast)
    \sin{2\rm \mathfrak{Re} \omega}\right \} \right],
\end{multline}
\vspace{-1em}
\begin{multline}\label{ImSet*Ret}
    {\rm Im}\left[S_{e\tau}^\ast R_{e\tau}\right]=\frac{\ln\frac{M_2}{M_1}}{M_1+M_2}{\rm Im}\left[S_{e \tau}^\ast \left \{(-X_{12}X_{32}^\ast+X_{13}X_{33}^\ast)\cos{2\rm \mathfrak{Re} \omega}+\right. \right.\\ \left. \left.
    +(X_{13}X_{32}^\ast+X_{12}X_{33}^\ast)
    \sin{2\rm \mathfrak{Re} \omega}\right \} \right],
\end{multline}
\vspace{-1em}
\begin{multline}\label{ImSmt*Rmt}
    {\rm Im}\left[S_{\mu\tau}^\ast R_{\mu\tau}\right]=\frac{\ln\frac{M_2}{M_1}}{M_1+M_2}{\rm Im}\left[S_{\mu\tau}^\ast \left \{(-X_{22}X_{32}^\ast+X_{23}X_{33}^\ast)\cos{2\rm \mathfrak{Re} \omega}+\right. \right.\\ \left. \left.
    +(X_{23}X_{32}^\ast+X_{22}X_{33}^\ast)
    \sin{2\rm \mathfrak{Re} \omega}\right \}\right].
\end{multline}
Expressions for the case of inverted ordering of active neutrino can be obtained from \eqref{ImSet*Ret}-\eqref{ImSmt*Rmt} by decreasing the
 last  index in $X$ matrix by one, e.g., $X_{32}\rightarrow X_{31}$.

Then, using explicit expressions for matrix $X_{\alpha \beta}$ \eqref{Xmatrix}, we can derive upper bounds for expressions \eqref{ImSem*Rem}--\eqref{ImSmt*Rmt}, see Appendix \ref{Append_Im_Lim}.

For the case of normal ordering of active neutrino masses, we get
\begin{equation}\label{ImS*R_NO}
     \left|{\rm Im}\left[S_{\alpha \beta}^\ast R_{\alpha \beta}\right]\right|\leq |S_{\alpha \beta}|\,\frac{\ln\frac{M_2}{M1}}{M_1+M_2} \frac{\sqrt{m_3^2+4m_2^2+8m_3 m_2}}{2 v^2}.
\end{equation}
This allows us to  obtain the following form for  \eqref{Sum_deltaa_FF*} for the case of normal ordering of active neutrino
masses
\begin{equation}\label{Deltaa_FF*_NO}
      \sum_{\alpha, I} \delta^\alpha |F_{\alpha I}|^2  \leq
    \frac{M_1^3 M_2^3}{2v^2(M_2+M_1)} \sqrt{m_3^2+4m_2^2+8m_3 m_2}
    \sum_{\alpha,\beta\neq\alpha} S_{\alpha \alpha} |S_{\alpha \beta}|.
\end{equation}
Then, for the case of  normal ordering of active neutrino
masses, we  obtain the following form  for the baryon asymmetry of the Universe
\begin{multline}\label{DeltaB_Form}
  \frac{n_B}{s}=7\cdot 10^{-4} \frac{\pi^{\frac32}\sin^3\phi}{384\cdot3^{\frac13}\Gamma(\frac56)}
    \frac{m^{\frac12}_{sol} m^{\frac32}_{atm} M_1^{\frac12} M_2^{\frac32} M_0^{\frac73}}{v^4 T_W (\Delta M^2_{32})^{\frac23}}  \dfrac{4}{f_1f_2(f_2^2 - f_1^2)}  \sum_{\alpha, I} \delta^\alpha |F_{\alpha I}|^2 
    \leq \\
    \frac{7 \cdot 10^{-4}\pi^{\frac32}\sin^3\phi}{384\cdot12^{\frac13}\Gamma(\frac56)} \frac{M_0^{\frac73} M^{\frac{11}{3}}\sqrt{m_3^2+4m_2^2+8m_3 m_2}}{T_W v^2 M_W^4 }\left(\frac{M}{\Delta M_{21}}\right)^{\frac{2}{3}} 
    \sum_{\alpha,\beta\neq\alpha}\hat S_{\alpha \alpha}|\hat S_{\alpha \beta}|.
\end{multline}
Here we take into account inequality \eqref{Deltaa_FF*_NO}, $M^2 m^{\frac12}_{sol} m^{\frac32}_{atm
}/(v^4f_1f_2(f_2^2 - f_1^2))\approx 1$,  $M_2\approx M_1 \approx M$ and
 $\Delta M_{21}^2=2M\Delta M_{21}$.  For the case of inverted ordering of active neutrino
masses, one needs to change the masses of active neutrinos $m_2$, $m_3$ \eqref{normalIO} to $m_1$, $m_2$ \eqref{invertedIO} in \eqref{ImS*R_NO} -- \eqref{DeltaB_Form}.

After substituting values of known parameters, one can get
\begin{equation}\label{limS}
    \frac{n_B}{s} \leq a\cdot 10^6 \left(\frac{M}{1\rm GeV}\right)^{11/3} \left(\frac{M}{\Delta M_{21}}\right)^{\frac{2}{3}}\sum_{\alpha,\beta\neq\alpha}\hat S_{\alpha \alpha}|\hat S_{\alpha \beta}|,
\end{equation}
where $a=4.3$ for the case of normal ordering and $a=9.9$ for the case of inverted ordering of active neutrino
masses. This relation allows us to set a lower limit on the combination of the matrix elements $S_{\alpha\beta}$.

Using results of Section 4 connecting elements of $S$ matrix with $U^2_{total}$ one can get
\begin{equation}\label{limU}
    \frac{n_B}{s} \leq b\cdot 10^4 \left(\frac{M}{1\rm GeV}\right)^{11/3} \left(\frac{M}{\Delta M_{21}}\right)^{\frac{2}{3}}U^4_{total},
\end{equation}
where $b=3.1$ for the case of normal ordering and $b=7.4$ for the case of inverted ordering of active neutrino
masses. This relation allows us to set a lower limit on $U^2_{total}$.

It should be noted that in \cite{Drewes:2021nqr,Ruchayskiy:2022,Granelli:2022eru} constraints from the lepton flavor violating process were also considered from the position of generating necessary lepton asymmetry for obtaining baryon asymmetry in the early Universe. However, only upper limits on these constraints were obtained, see the black line on left plot Fig.\ref{fig:coulomb_states2}.

Taking into account experimental constraints for $\hat R$ and $\hat S$ matrices from the Tabl.\ref{tab:my_label1} and Tabl.\ref{tab:my_label2}, one can see that the dominant contribution will arise from the following matrix elements 
\begin{equation}\label{factorS}
    \sum_{\alpha,\beta\neq\alpha}\hat S^{exp}_{\alpha \alpha} |\hat S^{exp}_{\alpha \beta}|\approx (\hat S^{exp}_{ee}+\hat  S^{exp}_{\mu\mu}+2\hat S^{exp}_{\tau\tau})|\hat  S^{exp}_{e\tau}|.
\end{equation}
Since the element $|\hat S_{e\tau}|$ is defined via diagonal elements  $|\hat  S_{e\tau}|=\sqrt{\hat  S_{ee} \hat  S_{\tau\tau}}$, we can conclude that research of the diagonal elements $\hat  S_{ee}$, $\hat S_{\mu\mu}$, $\hat S_{ee}$  is a priority task.
The relation \eqref{factorS} allows us to get the following inequalities.

For the case of  normal ordering of active neutrino
masses,
\begin{equation}\label{DB1}
    \frac{n_B}{s} \leq 4.6\left(\frac{M}{\Delta M_{21}}\right)^{\frac{2}{3}}(M/1 {\rm GeV})^\frac{11}{3}.
\end{equation}

For the case of inverted ordering of active neutrino
masses,
\begin{equation}\label{DB2}
    \frac{n_B}{s} \leq 10.4\left(\frac{M}{\Delta M_{21}}\right)^{\frac{2}{3}}(M/1 {\rm GeV})^\frac{11}{3}.
\end{equation}

However, these inequalities have no practical use. The left side is very small (${n_B}/{s}\sim 10^{-10}$), whereas the right side is significantly greater than 1 because ${M}/{\Delta M_{21}}\gg 1$ and $M \gtrsim 1$ GeV. This can be explained by assuming that the actual values of elements of $\hat R$ and $\hat S$ matrices are many orders of magnitude less than the experimental constraints given in the Tabl.\ref{tab:my_label1}. 

It should be noted that the expression \eqref{g11} used for baryonic asymmetry suffers from divergence in the case of exactly degenerate limit $M_1 \rightarrow M_2$. So, obtained estimates \eqref{DB1}, \eqref{DB2} can be improved by taking into account the finite lifetime of the HNL and modifying the denominator $\Delta M^2_{21}\rightarrow \Delta M^2_{21}+A$, where $A\sim M_I \Gamma_I$ \cite{Klaric:2021cpi}.

\section{Conclusions}

This paper focuses on the neutrino extension of the SM by adding two heavy right-handed neutrinos with masses significantly above the electroweak scale. The direct observation of these heavy neutrinos is a challenging test. 
However, these particles can generate charged lepton flavor violation (cLFV) processes via effective dimension-6 operators. 
Non-observation of cLFV processes allows us to impose restrictions on HNL parameters.

We have analytically derived relationships between observable parameters of the neutrino extension of the SM with two heavy RH neutrinos of different masses, and observable parameters of massive active neutrinos, see \eqref{SandR} and \eqref{SRrelations}.  To fulfil this relationship, the masses of the active neutrinos must be extremely small compared to the masses of the RH neutrinos. Additionally, the condition \eqref{BigOmegaAssum}, which is relevant for the current experimental search for HNL, must be satisfied. 
We derive analytically that Schwarz inequalities $|S_{\alpha \beta}|^2 \leq S_{\alpha \alpha} S_{\beta \beta} $ and $|R_{\alpha \beta}|^2 \leq R_{\alpha \alpha} R_{\beta \beta} $ saturate only when $e^{\mathfrak{Im}\omega}\gg 1$. The independence of the obtained result from the difference in mass of sterile neutrinos is very important. This is because the commonly used case of almost degenerate values of the masses of RH neutrinos is still only a theoretical assumption. In particular, this assumption is used for the description of the baryon asymmetry of the Universe in the  Minimal Neutrino Modification of the SM ($\nu$MSM) \cite{Asaka:2005,Asaka_2:2005}. However
the baryon asymmetry of the Universe can also be generated in models with different RH neutrino masses, see e.g. \cite{Drewes:2012ma}.

It should be noted that previous similar relations $|S_{\alpha \beta}|^2 \leq S_{\alpha \alpha} S_{\beta \beta} $ and $|R_{\alpha \beta}|^2 \leq R_{\alpha \alpha} R_{\beta \beta} $ in the form of saturated Schwarz inequality were obtained only for the partial case of massless active neutrinos or 
degenerate masses of heavy sterile neutrinos \cite{Coy:2018bxr,Blennow:2023mqx}. Obtained in our paper relations $|S_{\alpha \beta}|^2 = S_{\alpha \alpha} S_{\beta \beta} $ and $|R_{\alpha \beta}|^2 = R_{\alpha \alpha} R_{\beta \beta} $ that are valid for actual parameters of the neutrino modification of the SM decrease number of independent parameters for experimental search. These relations do not depend on the mass scale of the RH neutrino. In fact, knowing of experimental constraints on diagonal elements of $S$- and $R$-matrices allows us to find constraints on non-diagonal elements.

Assuming that the restrictions \eqref{SRrelations} are satisfied with sufficient accuracy for the condition \eqref{BigOmegaAssum} above the see-saw line we find constraints on the tau-flavours from much stronger constraints on muon and electron flavors. It turns out that our predictions provide better constraints than the experimental data for $\hat S_{e\tau}$ and $\hat S_{\mu\tau}$ elements by an order of magnitude, see Tabl.\ref{tab:my_label2}.

Using experimental constraints, we have constructed graphical regions of allowed values for parameters $U^2_{tot}={\displaystyle \sum_{\alpha,I}} |\Theta_{\alpha I}|^2$ and $M$ (degenerate mass of the two heavy neutrinos $M_1=M_2=M$), which are agree quite well with the results
 of a more complicated
computations \cite{Ruchayskiy:2022}, see Fig.\ref{fig:coulomb_states} and Fig.\ref{fig:coulomb_states2}.

We also consider the obtained general relations that connect the elements of the $\hat S$ and $\hat R$ matrices with the observed parameters of active neutrinos and the parameters of the SM expansion of two RH neutrinos with different masses and arbitrary values of $\cosh2\mathfrak{Im}\omega$ \eqref{UtotDef} to be useful for further analytical studies, see Appendix \ref{Append_Sab_Rab}.


We obtained an expression for baryon asymmetry of the Universe, using experimentally observable parameters (elements of $\hat S$ and $\hat R$ matrices), see \eqref{DeltaB_Form}.  We found that, to a first approximation \eqref{BigOmegaAssum}, the baryon asymmetry of the Universe is zero. The non-zero baryon asymmetry of the Universe arises only when the subtle effect is taken into account when $\cosh2\mathfrak{Im}\omega \neq \sinh2\mathfrak{Im}\omega$.
We get a lower limit on the combination of the matrix elements $S_{\alpha\beta}$ \eqref{limS} or on the $U^2_{total}$ \label{limU} depending on mass $M$ and $\Delta M$ of the RH neutrinos. 
We have shown that the upper (from experiments at accelerators) and lower (from baryon asymmetry) bounds on the observable parameters of the neutrino extension of the SM differ by many orders of magnitude. This implies, if we believe that baryon asymmetry of the Universe is generated by heavy neutrinos of the $\nu$MSM model, that the actual values of elements of $\hat R$, $\hat S$ matrices are many orders of magnitude lower than experimental constraints given in Tabl.\ref{tab:my_label1}.   

The dominant contribution to the baryon asymmetry of the Universe is given by the diagonal elements $\hat  S_{ee}$, $\hat S_{\mu\mu}$, $\hat S_{ee}$. Consequently, investigating these elements is a priority task for experimental physics that
 requires high-precision measurements of electroweak reactions, see e.g. \cite{ALEPH:2005ab,Pich:2013lsa}. Even without concerning baryon asymmetry of the Universe, the diagonal elements $\hat  S_{\alpha\alpha}$ are important as they determine the off-diagonal elements of $\hat S$ and $\hat R$ matrices according to relations \eqref{SandR}, \eqref{SRrelations}.

\vspace{5mm}

\centerline{\bf Acknowledgements}
\vspace{5mm}

The authors thank I. Timiryasov for fruitful discussions.  The work of V.G. and O.Kh. was supported by the National Research Foundation of Ukraine under project No. 2023.03/0149.

 \newpage

\appendix 

\section{Appendix: explicit relations for $S_{\alpha \beta}$ and  $R_{\alpha \beta}$ }\label{Append_Sab_Rab}

Expressions for the observable matrix elements $S_{\alpha\beta}$ \eqref{Sab} for normal and inverted ordering of active neutrino masses in the case of extension of the SM by two RH neutrinos for arbitrary values of parameters of the model.

 Normal ordering:
\begin{multline}\label{SeeFullformNO}
     S_{ee}= 
     \frac{M_2-M_1}{2 M_1 M_2} \bigl[(|X_{12}|^2-|X_{13}|^2)
  \cos{2 \mathfrak{Re}\omega}-(X_{12}X_{13}^\ast + X_{13}X_{12}^\ast)\sin{2 \mathfrak{Re}\omega}\bigr]\\ 
+\frac{M_1+M_2}{2 M_1 M_2} \bigl[(|X_{12}|^2+|X_{13}|^2)
  \cosh{2 \mathfrak{Im}\omega}+i(X_{12}X_{13}^\ast-X_{13}X_{12}^\ast)\sinh{2 \mathfrak{Im}\omega}\bigr],
\end{multline}
\vspace{-1em}
\begin{multline}\label{SmumuFullformNO}
     S_{\mu \mu}=\frac{M_2-M_1}{2 M_1 M_2} \bigl[(|X_{22}|^2-|X_{23}|^2)
  \cos{2 \mathfrak{Re}\omega}-(X_{22}X_{23}^\ast + X_{23}X_{22}^\ast)\sin{2 \mathfrak{Re}\omega}\bigr]\\
  + \frac{M_1+M_2}{2 M_1 M_2} \bigl[(|X_{22}|^2+|X_{23}|^2)
  \cosh{2 \mathfrak{Im}\omega}+i(X_{22}X_{23}^\ast-X_{23}X_{22}^\ast)\sinh{2 \mathfrak{Im}\omega}\bigr],
\end{multline}
\vspace{-1em}
\begin{multline}\label{StautauFullformNO}
     S_{\tau \tau}=\frac{M_2-M_1}{2 M_1 M_2} \bigl[(|X_{32}|^2-|X_{33}|^2)
  \cos{2 \mathfrak{Re}\omega}-(X_{32}X_{33}^\ast + X_{33}X_{32}^\ast)\sin{2 \mathfrak{Re}\omega}\bigr]\\
  + \frac{M_1+M_2}{2 M_1 M_2}\bigl[(|X_{32}|^2+|X_{33}|^2)
  \cosh{2 \mathfrak{Im}\omega}+i(X_{32}X_{33}^\ast-X_{33}X_{32}^\ast)\sinh{2 \mathfrak{Im}\omega}\bigr],
\end{multline}
\vspace{-1em}
\begin{multline}\label{SemuFullformNO}
     S_{e \mu}= \frac{M_2-M_1}{2 M_1 M_2} \bigl[(X_{12}X_{22}^\ast-X_{13}X_{23}^\ast)
  \cos{2 \mathfrak{Re}\omega}-(X_{12}X_{23}^\ast + X_{13}X_{22}^\ast)\sin{2 \mathfrak{Re}\omega}\bigr]\\
 + \frac{M_1+M_2}{2 M_1 M_2} \bigl[(X_{12}X_{22}^\ast+X_{13}X_{23}^\ast)
  \cosh{2 \mathfrak{Im}\omega}+i(X_{12}X_{23}^\ast-X_{13}X_{22}^\ast)\sinh{2 \mathfrak{Im}\omega}\bigr],
\end{multline}
\vspace{-1em}
\begin{multline}\label{SetauFullformNO}
     S_{e \tau}= \frac{M_2-M_1}{2 M_1 M_2} \bigl[(X_{22}X_{32}^\ast-X_{23}X_{33}^\ast)
  \cos{2 \mathfrak{Re}\omega}-(X_{22}X_{33}^\ast + X_{23}X_{32}^\ast)\sin{2 \mathfrak{Re}\omega}\bigr] \\
 + \frac{M_1+M_2}{2 M_1 M_2} \bigl[(X_{12}X_{32}^\ast+X_{13}X_{33}^\ast)
  \cosh{2 \mathfrak{Im}\omega}+i(X_{22}X_{33}^\ast-X_{23}X_{32}^\ast)\sinh{2 \mathfrak{Im}\omega}\bigr],
\end{multline}
\vspace{-1em}
\begin{multline}\label{SmutauFullformNO}
     S_{\mu \tau}= \frac{M_2-M_1}{2 M_1 M_2} \bigl[(X_{12}X_{32}^\ast-X_{13}X_{33}^\ast)
  \cos{2 \mathfrak{Re}\omega}-(X_{12}X_{33}^\ast + X_{13}X_{32}^\ast)\sin{2 \mathfrak{Re}\omega}\bigr] \\
  +\frac{M_1+M_2}{2 M_1 M_2} \bigl[(X_{12}X_{32}^\ast+X_{13}X_{33}^\ast)
  \cosh{2 \mathfrak{Im}\omega}+i(X_{12}X_{33}^\ast-X_{13}X_{32}^\ast)\sinh{2 \mathfrak{Im}\omega}\bigr],
\end{multline}
\vspace{-1em}
\begin{multline}\label{ReeFullformNO}
     R_{ee}= 
     M_2^{-1}\ln{\frac{M_2}{M_W}}\, \left|X_{13}\cosh( -i \omega)+i X_{12}\sinh( -i \omega)\right|^2 \\
    + M_1^{-1}\ln{\frac{M_1}{M_W}}\, \left|X_{12}\cosh{(-i\omega)}-i X_{13}\sinh{(-i\omega)}\right|^2,
\end{multline}
\vspace{-1em}
\begin{multline}\label{RmumuFullformNO}
     R_{\mu \mu}= 
     M_2^{-1}\ln{\frac{M_2}{M_W}}\,\left|X_{23}\cosh{( -i \omega)}+i X_{22}\sinh{( -i \omega)}\right|^2\\
    + M_1^{-1}\ln{\frac{M_1}{M_W}}\,\left|X_{22}\cosh{( -i \omega)}-i X_{23}\sinh{( -i \omega)}\right|^2,
\end{multline}
\vspace{-1em}
\begin{multline}\label{RtautauFullformNO}
     R_{\tau \tau}= 
     M_2^{-1}\ln{\frac{M_2}{M_W}}\,\left|X_{33}\cosh{( -i \omega)}+i X_{32}\sinh{( -i \omega)}\right|^2\\
    + M_1^{-1}\ln{\frac{M_1}{M_W}}\,\left|X_{32}\cosh{( -i \omega)}-i X_{33}\sinh{( -i \omega)}\right|^2,
\end{multline}
\vspace{-1em}
\begin{multline}\label{RemuFullformNO}
     R_{e\mu}= 
     M_2^{-1}\ln{\frac{M_2}{M_W}}\,(X_{13}\cosh{( -i \omega)}+i X_{12}\sinh{( -i \omega)})
      (X_{23}^\ast\cosh{( i \omega^*)}-i X_{22}^\ast\sinh{( i \omega^*)})\\
    + M_1^{-1}\ln{\frac{M_1}{M_W}}\,(X_{12}\cosh{( -i \omega)}-i X_{13}\sinh{( -i \omega)})
      (X_{22}^\ast\cosh{( i \omega^*)}+i X_{23}^\ast\sinh{( i \omega^*)}),
\end{multline}
\vspace{-1em}
\begin{multline}\label{RetauFullformNO}
     R_{e\tau}= 
     M_2^{-1}\ln{\frac{M_2}{M_W}}\,(X_{13}\cosh{( -i \omega)}+i X_{12}\sinh{( -i \omega)})
      (X_{33}^\ast\cosh{( i \omega^*)}-i X_{32}^\ast\sinh{( i \omega^*)})\\
    + M_1^{-1}\ln{\frac{M_1}{M_W}}\,(X_{12}\cosh{( -i \omega)}-i X_{13}\sinh{( -i \omega)}) (X_{32}^\ast\cosh{( i \omega^*)}+i X_{33}^\ast\sinh{( i \omega^*)}),
\end{multline}
\vspace{-1em}
\begin{multline}\label{RmutauFullformNO}
     R_{\mu\tau}= 
     M_2^{-1}\ln{\frac{M_2}{M_W}}\,(X_{23}\cosh{( -i \omega)}+i X_{22}\sinh{( -i \omega)}) (X_{33}^\ast\cosh{( i \omega^*)}-i X_{32}^\ast\sinh{( i \omega^*)})\\
   + M_1^{-1}\ln{\frac{M_1}{M_W}}\,(X_{22}\cosh{( -i \omega)}-i X_{23}\sinh{( -i \omega)}) (X_{32}^\ast\cosh{( i \omega^*)}+i X_{33}^\ast\sinh{( i \omega^*)}).
\end{multline}

 Inverted ordering:
\begin{multline}\label{SeeFullformIO}
     S_{ee}= \frac{M_2-M_1}{2 M_1 M_2} \bigl[(|X_{11}|^2-|X_{12}|^2)
  \cos{2 \mathfrak{Re}\omega}-(X_{11}X_{12}^\ast + X_{12}X_{11}^\ast)\sin{2 \mathfrak{Re}\omega}\bigr]\\
 + \frac{M_1+M_2}{2 M_1 M_2} \bigl[(|X_{11}|^2+|X_{12}|^2)
  \cosh{2 \mathfrak{Im}\omega}+i(X_{11}X_{12}^\ast-X_{12}X_{11}^\ast)\sinh{2 \mathfrak{Im}\omega}\bigr],
\end{multline}
\vspace{-1em}
\begin{multline}\label{SmumuFullformIO}
     S_{\mu \mu}= \frac{M_2-M_1}{2 M_1 M_2} \bigl[(|X_{21}|^2-|X_{22}|^2)
  \cos{2 \mathfrak{Re}\omega}-(X_{21}X_{22}^\ast + X_{22}X_{21}^\ast)\sin{2 \mathfrak{Re}\omega}\bigr]\\
 + \frac{M_1+M_2}{2 M_1 M_2} \bigl[(|X_{21}|^2+|X_{22}|^2)
  \cosh{2 \mathfrak{Im}\omega}+i(X_{21}X_{22}^\ast-X_{22}X_{21}^\ast)\sinh{2 \mathfrak{Im}\omega}\bigr],
\end{multline}
\vspace{-1em}
\begin{multline}\label{StautauFullformIO}
     S_{\tau \tau}= \frac{M_2-M_1}{2 M_1 M_2} \bigl[(|X_{31}|^2-|X_{32}|^2)
  \cos{2 \mathfrak{Re}\omega}-(X_{31}X_{32}^\ast + X_{32}X_{31}^\ast)\sin{2 \mathfrak{Re}\omega}\bigr]\\
 + \frac{M_1+M_2}{2 M_1 M_2} \bigl[(|X_{31}|^2+|X_{32}|^2)
  \cosh{2 \mathfrak{Im}\omega}+i(X_{31}X_{32}^\ast-X_{32}X_{31}^\ast)\sinh{2 \mathfrak{Im}\omega}\bigr],
\end{multline}
\vspace{-1em}
\begin{multline}\label{SemuFullformIO}
     S_{e \mu}= \frac{M_2-M_1}{2 M_1 M_2} \bigl[(X_{11}X_{21}^\ast-X_{12}X_{22}^\ast)
  \cos{2 \mathfrak{Re}\omega}-(X_{11}X_{22}^\ast + X_{12}X_{21}^\ast)\sin{2 \mathfrak{Re}\omega}\bigr]\\
 + \frac{M_1+M_2}{2 M_1 M_2} \bigl[(X_{11}X_{21}^\ast+X_{12}X_{22}^\ast)
  \cosh{2 \mathfrak{Im}\omega}+i(X_{11}X_{22}^\ast-X_{12}X_{21}^\ast)\sinh{2 \mathfrak{Im}\omega}\bigr],
\end{multline}
\vspace{-1em}
\begin{multline}\label{SetauFullformIO}
     S_{e \tau}= \frac{M_2-M_1}{2 M_1 M_2} \bigl[(X_{21}X_{31}^\ast-X_{22}X_{32}^\ast)
  \cos{2 \mathfrak{Re}\omega}-(X_{21}X_{32}^\ast + X_{22}X_{31}^\ast)\sin{2 \mathfrak{Re}\omega}\bigr]\\
 + \frac{M_1+M_2}{2 M_1 M_2} \bigl[(X_{11}X_{31}^\ast+X_{12}X_{32}^\ast)
  \cosh{2 \mathfrak{Im}\omega}+i(X_{21}X_{32}^\ast-X_{22}X_{31}^\ast)\sinh{2 \mathfrak{Im}\omega}\bigr],
\end{multline}
\vspace{-1em}
\begin{multline}\label{SmutauFullformIO}
     S_{\mu \tau}= \frac{M_2-M_1}{2 M_1 M_2} \bigl[(X_{11}X_{31}^\ast-X_{12}X_{32}^\ast)
  \cos{2 \mathfrak{Re}\omega}-(X_{11}X_{32}^\ast + X_{12}X_{31}^\ast)\sin{2 \mathfrak{Re}\omega}\bigr]\\
 + \frac{M_1+M_2}{2 M_1 M_2} \bigl[(X_{11}X_{31}^\ast+X_{12}X_{32}^\ast)
  \cosh{2 \mathfrak{Im}\omega}+i(X_{11}X_{32}^\ast-X_{12}X_{31}^\ast)\sinh{2 \mathfrak{Im}\omega}\bigr],
\end{multline}
\vspace{-1em}
\begin{multline}\label{ReeFullformIO}
     R_{ee}= 
     M_2^{-1}\ln{\frac{M_2}{M_W}}\,\left|X_{12}\cosh{( -i \omega)}+i X_{11}\sinh{( -i \omega)}\right|^2\\
    + M_1^{-1}\ln{\frac{M_1}{M_W}}\,\left|X_{11}\cosh{( -i \omega)}-i X_{12}\sinh{( -i \omega)}\right|^2,
\end{multline}
\vspace{-1em}
\begin{multline}\label{RmumuFullformIO}
     R_{\mu \mu}= 
     M_2^{-1}\ln{\frac{M_2}{M_W}}\,\left|X_{22}\cosh{( -i \omega)}+i X_{21}\sinh{( -i \omega)}\right|^2\\
    + M_1^{-1}\ln{\frac{M_1}{M_W}}\,\left|X_{21}\cosh{( -i \omega)}-i X_{22}\sinh{( -i \omega)}\right|^2,
\end{multline}
\vspace{-1em}
\begin{multline}\label{RtautauFullformIO}
     R_{\tau \tau}= 
     M_2^{-1}\ln{\frac{M_2}{M_W}}\,\left|X_{32}\cosh{( -i \omega)}+i X_{31}\sinh{( -i \omega)}\right|^2\\
    + M_1^{-1}\ln{\frac{M_1}{M_W}}\,\left|X_{31}\cosh{( -i \omega)}-i X_{32}\sinh{( -i \omega)}\right|^2,
\end{multline}
\vspace{-1em}
\begin{multline}\label{RemuFullformIO}
     R_{e\mu}= 
    M_2^{-1}\ln{\frac{M_2}{M_W}}\,(X_{12}\cosh{( -i \omega)}+i X_{11}\sinh{( -i \omega)}) (X_{22}^\ast\cosh{( i \omega^*)}-i X_{21}^\ast\sinh{( i \omega^*)})\\
    + M_1^{-1}\ln{\frac{M_1}{M_W}}\,(X_{11}\cosh{( -i \omega)}-i X_{12}\sinh{( -i \omega)}) (X_{21}^\ast\cosh{( i \omega^*)}+i X_{22}^\ast\sinh{( i \omega^*)}),
\end{multline}
\vspace{-1em}
\begin{multline}\label{RetauFullformIO}
     R_{e\tau}= 
     M_2^{-1}\ln{\frac{M_2}{M_W}}\,(X_{12}\cosh{( -i \omega)}+i X_{11}\sinh{( -i \omega)}) (X_{32}^\ast\cosh{( i \omega^*)}-i X_{31}^\ast\sinh{( i \omega^*)})\\
    + M_1^{-1}\ln{\frac{M_1}{M_W}}\,(X_{11}\cosh{( -i \omega)}-i X_{12}\sinh{( -i \omega)}) (X_{31}^\ast\cosh{( i \omega^*)}+i X_{32}^\ast\sinh{( i \omega^*)}),
\end{multline}
\vspace{-1em}
\begin{multline}\label{RmutauFullformIO}
     R_{\mu\tau}= 
    M_2^{-1}\ln{\frac{M_2}{M_W}}\,(X_{22}\cosh{( -i \omega)}+i X_{21}\sinh{( -i \omega)}) (X_{32}^\ast\cosh{( i \omega^*)}-i X_{31}^\ast\sinh{( i \omega^*)})\\
   + M_1^{-1}\ln{\frac{M_1}{M_W}}\,(X_{21}\cosh{( -i \omega)}-i X_{22}\sinh{( -i \omega)}) (X_{31}^\ast\cosh{( i \omega^*)}+i X_{32}^\ast\sinh{( i \omega^*)}).
\end{multline}

\section{Appendix: upper limit estimate for $\left|{\rm Im}\left[S_{\alpha \beta}^\ast R_{\alpha \beta}\right]\right|$ }\label{Append_Im_Lim}
For the case of the normal ordering of active neutrinos, one can rewrite the relation
\eqref{ImSem*Rem} for ${\rm Im}\left[S_{e\mu}^\ast R_{e\mu}\right]$ in the following form
\begin{equation}
    {\rm Im}\left[S_{e\mu}^\ast R_{e\mu}\right]=\frac{\ln\frac{M_2}{M_1}}{M_1+M_2}{\rm Im}\left[S_{e \mu}^\ast K_{e \mu}\right],
\end{equation}
where
\begin{equation} \label{K_emu}
    K_{e\mu}=(-X_{12}X_{22}^\ast+X_{13}X_{23}^\ast)\cos{2\rm \mathfrak{Re} \omega}+(X_{13}X_{22}^\ast+X_{12}X_{23}^\ast)
    \sin{2\rm \mathfrak{Re} \omega}.
\end{equation}

We can estimate an upper limit for this expression as
\begin{equation} \label{ImLim_emu }
     \left|{\rm Im}\left[S_{e\mu}^\ast R_{e\mu}\right]\right|\leq\frac{\ln\frac{M_2}{M_1}}{M_1+M_2} |S_{e \mu}| |K_{e\mu}|.
\end{equation}

Using expressions for matrix $X$ \eqref{Xmatrix}, we can rewrite \eqref{K_emu} in terms of mixing angles and active neutrino masses
\begin{multline}
    \left|K_{e \mu} \right|=\frac{1}{v^2} \left| \cos{2\rm \mathfrak{Re} \omega} \cos \theta_{13}\times\right. \\ \left(-m_2 \cos \theta_{12} \cos \theta_{23} \sin \theta_{12} + e^{-i\delta} \left(m_3 + m_2 \sin^2 \theta_{12}\right)\sin \theta_{13} \sin \theta_{23}\right)\\
    + e^{-i\left(a_2+4\delta\right)/2}
    \sin{2\rm \mathfrak{Re} \omega}\sqrt{m_2 m_3}\times \\ \left. \left(e^{i\delta}cos \theta_{12} \cos \theta_{23} \sin \theta_{13}+\sin \theta_{12}\sin \theta_{23}\left(e^{i\left(a_2+2\delta\right)}\cos^2 \theta_{13}-\sin^2\theta_{13}\right)\right)\right|.
\end{multline}
Than we can estimate an upper limit for $K_{e \mu}$ using this form
\begin{multline}
    \left|K_{e \mu} \right|\leq\frac{1}{v^2} \left \{m_3\left| \cos{2\rm \mathfrak{Re} \omega} \cos \theta_{13}\sin \theta_{13} \sin \theta_{23}\right|\right.\\
    +\left.m_2\left|\cos{2\rm \mathfrak{Re} \omega} \cos \theta_{13}\left(- \cos \theta_{12} \cos \theta_{23} \sin \theta_{12} + e^{-i\delta} \sin^2 \theta_{12}\sin \theta_{13} \sin \theta_{23}\right)\right|\right.\\
    +\left.\sqrt{m_2 m_3} \left|
    \sin{2\rm \mathfrak{Re} \omega}\left(e^{i\delta}cos \theta_{12} \cos \theta_{23} \sin \theta_{13}+\sin \theta_{12}\sin \theta_{23}\left(e^{i\left(a_2+2\delta\right)}\cos^2 \theta_{13}-\sin^2\theta_{13}\right)\right)\right| \right\}.
\end{multline}

It is easier to consider the terms with different masses separately.
For the term near $m_3$, we can obtain
\begin{equation}
    \left| \cos{2\rm \mathfrak{Re} \omega} \cos \theta_{13}\sin \theta_{13} \sin \theta_{23}\right|=\frac{1}{2}\left| \cos{2\rm \mathfrak{Re} \omega} \sin 2\theta_{13} \sin \theta_{23}\right|\leq \frac{1}{2} \left| \cos{2\rm \mathfrak{Re} \omega}\right|.
\end{equation}
For the term near $m_2$, we can obtain
\begin{multline}
    \left|\cos{2\rm \mathfrak{Re} \omega} \cos \theta_{13}\left(- \cos \theta_{12} \cos \theta_{23} \sin \theta_{12} + e^{-i\delta} \sin^2 \theta_{12}\sin \theta_{13} \sin \theta_{23}\right)\right|\\
    \leq \left|\cos{2\rm \mathfrak{Re} \omega} \cos \theta_{13}\sin \theta_{12}\right|\leq \left|\cos{2\rm \mathfrak{Re} \omega} \right|.
\end{multline}
For the term near $\sqrt{m_2 m_3}$, we can obtain
\begin{multline}
    \left|\sin{2\rm \mathfrak{Re} \omega}\left(e^{i\delta}cos \theta_{12} \cos \theta_{23} \sin \theta_{13}+\sin \theta_{12}\sin \theta_{23}\left(e^{i\left(a_2+2\delta\right)}\cos^2 \theta_{13}-\sin^2\theta_{13}\right)\right)\right|\\
    \leq \left|\sin{2\rm \mathfrak{Re} \omega}\left(e^{i\delta}cos \theta_{12} \cos \theta_{23} \sin \theta_{13}+\sin \theta_{12}\sin \theta_{23}\right)\right|\leq \left|\sin{2\rm \mathfrak{Re} \omega}\right|.
\end{multline}
Combining all the terms we get
\begin{multline}
    \left|K_{e \mu} \right|\leq\frac{1}{2v^2} \left \{m_3\left| \cos{2\rm \mathfrak{Re} \omega} \right| + 2 m_2\left| \cos{2\rm \mathfrak{Re} \omega} \right| +2 \sqrt{m_2 m_3} \left| \sin{2\rm \mathfrak{Re} \omega} \right| \right\}\leq \\
    \frac{1}{2v^2}  \sqrt{m_3^2 + 4 m_2^2 + 8 m_3 m_2}.
\end{multline}

Using equation \eqref{ImLim_emu } we obtain
\begin{equation}
     \left|{\rm Im}\left[S_{e\mu}^\ast R_{e\mu}\right]\right|\leq |S_{e \mu}| \frac{\ln\frac{M_2}{M_1}}{M_1+M_2} \frac{1}{2v^2} \sqrt{m_3^2 + 4 m_2^2 + 8 m_3 m_2}.
\end{equation}
A similar expression can also be obtained for $\left|{\rm Im}\left[S_{e\tau}^\ast R_{e\tau}\right]\right|, \left|{\rm Im}\left[S_{\mu\tau}^\ast R_{\mu\tau}\right]\right|$ via the same method.

In general, for normal ordering  of active neutrino masses, we obtain
\begin{equation}\label{ImLim_NO}
     \left|{\rm Im}\left[S_{\alpha \beta}^\ast R_{\alpha \beta}\right]\right|\leq |S_{\alpha \beta}| \frac{\ln\frac{M_2}{M_1}}{M_1+M_2} \frac{1}{2v^2} \sqrt{m_3^2 + 4 m_2^2 + 8 m_3 m_2}.
\end{equation}

We obtain similar expressions for inverted ordering of active neutrino masses
\begin{equation}\label{ImLim_IO}
     \left|{\rm Im}\left[S_{\alpha \beta}^\ast R_{\alpha \beta}\right]\right|\leq |S_{\alpha \beta}| \frac{\ln\frac{M_2}{M_1}}{M_1+M_2} \frac{1}{2v^2} \sqrt{m_2^2 + 4 m_1^2 + 8 m_2 m_1}.
\end{equation}

 \newpage

\bibliographystyle{JHEP}
\bibliography{Bibl}

\providecommand{\href}[2]{#2}\begingroup\raggedright\begin{thebibliography}{10}

\bibitem{Cottingham:2007zz}
W.N.~Cottingham and D.A.~Greenwood, \emph{{An Introduction to the Standard Model of Particle Physics}}, Cambridge University Press (7, 2023), \href{https://doi.org/10.1017/9781009401685}{10.1017/9781009401685}.

\bibitem{Bilenky:1987ty}
S.M.~Bilenky and S.T.~Petcov, \emph{{Massive Neutrinos and Neutrino Oscillations}}, \href{https://doi.org/10.1103/RevModPhys.59.671}{\emph{Rev. Mod. Phys.} {\bfseries 59} (1987) 671}.

\bibitem{Strumia:2006db}
A.~Strumia and F.~Vissani, \emph{{Neutrino masses and mixings and...}},  \href{https://arxiv.org/abs/hep-ph/0606054}{{\ttfamily hep-ph/0606054}}.

\bibitem{deSalas:2017kay}
P.F.~de~Salas, D.V.~Forero, C.A.~Ternes, M.~Tortola and J.W.F.~Valle, \emph{{Status of neutrino oscillations 2018: 3$\sigma$ hint for normal mass ordering and improved CP sensitivity}}, \href{https://doi.org/10.1016/j.physletb.2018.06.019}{\emph{Phys. Lett. B} {\bfseries 782} (2018) 633} [\href{https://arxiv.org/abs/1708.01186}{{\ttfamily 1708.01186}}].

\bibitem{Peebles:2013hla}
P.J.E.~Peebles, \emph{{Dark Matter}}, \href{https://doi.org/10.1073/pnas.1308786111}{\emph{Proc. Nat. Acad. Sci.} {\bfseries 112} (2015) 2246} [\href{https://arxiv.org/abs/1305.6859}{{\ttfamily 1305.6859}}].

\bibitem{Lukovic:2014vma}
V.~Lukovic, P.~Cabella and N.~Vittorio, \emph{{Dark matter in cosmology}}, \href{https://doi.org/10.1142/S0217751X14430015}{\emph{Int. J. Mod. Phys. A} {\bfseries 29} (2014) 1443001} [\href{https://arxiv.org/abs/1411.3556}{{\ttfamily 1411.3556}}].

\bibitem{Bertone:2016nfn}
G.~Bertone and D.~Hooper, \emph{{History of dark matter}}, \href{https://doi.org/10.1103/RevModPhys.90.045002}{\emph{Rev. Mod. Phys.} {\bfseries 90} (2018) 045002} [\href{https://arxiv.org/abs/1605.04909}{{\ttfamily 1605.04909}}].

\bibitem{Steigman:1976ev}
G.~Steigman, \emph{{Observational tests of antimatter cosmologies}}, \href{https://doi.org/10.1146/annurev.aa.14.090176.002011}{\emph{Ann. Rev. Astron. Astrophys.} {\bfseries 14} (1976) 339}.

\bibitem{Riotto:1999yt}
A.~Riotto and M.~Trodden, \emph{{Recent progress in baryogenesis}}, \href{https://doi.org/10.1146/annurev.nucl.49.1.35}{\emph{Ann. Rev. Nucl. Part. Sci.} {\bfseries 49} (1999) 35} [\href{https://arxiv.org/abs/hep-ph/9901362}{{\ttfamily hep-ph/9901362}}].

\bibitem{Canetti:2012zc}
L.~Canetti, M.~Drewes and M.~Shaposhnikov, \emph{{Matter and Antimatter in the Universe}}, \href{https://doi.org/10.1088/1367-2630/14/9/095012}{\emph{New J. Phys.} {\bfseries 14} (2012) 095012} [\href{https://arxiv.org/abs/1204.4186}{{\ttfamily 1204.4186}}].

\bibitem{Beacham:2020}
J.~Beacham et~al., \emph{{Physics Beyond Colliders at CERN: Beyond the Standard Model Working Group Report}}, \href{https://doi.org/10.1088/1361-6471/ab4cd2}{\emph{J. Phys. G} {\bfseries 47} (2020) 010501} [\href{https://arxiv.org/abs/1901.09966}{{\ttfamily 1901.09966}}].

\bibitem{Patt:2006fw}
B.~Patt and F.~Wilczek, \emph{{Higgs-field portal into hidden sectors}},  \href{https://arxiv.org/abs/hep-ph/0605188}{{\ttfamily hep-ph/0605188}}.

\bibitem{Bezrukov:2009yw}
F.~Bezrukov and D.~Gorbunov, \emph{{Light inflaton Hunter's Guide}}, \href{https://doi.org/10.1007/JHEP05(2010)010}{\emph{JHEP} {\bfseries 05} (2010) 010} [\href{https://arxiv.org/abs/0912.0390}{{\ttfamily 0912.0390}}].

\bibitem{Boiarska:2019jym}
I.~Boiarska, K.~Bondarenko, A.~Boyarsky, V.~Gorkavenko, M.~Ovchynnikov and A.~Sokolenko, \emph{{Phenomenology of GeV-scale scalar portal}}, \href{https://doi.org/10.1007/JHEP11(2019)162}{\emph{JHEP} {\bfseries 11} (2019) 162} [\href{https://arxiv.org/abs/1904.10447}{{\ttfamily 1904.10447}}].

\bibitem{Okun:1982xi}
L.B.~Okun, \emph{{LIMITS OF ELECTRODYNAMICS: PARAPHOTONS?}}, {\emph{Sov. Phys. JETP} {\bfseries 56} (1982) 502}.

\bibitem{Holdom:1985ag}
B.~Holdom, \emph{{Two U(1)'s and Epsilon Charge Shifts}}, \href{https://doi.org/10.1016/0370-2693(86)91377-8}{\emph{Phys. Lett. B} {\bfseries 166} (1986) 196}.

\bibitem{Langacker:2008yv}
P.~Langacker, \emph{{The Physics of Heavy $Z^\prime$ Gauge Bosons}}, \href{https://doi.org/10.1103/RevModPhys.81.1199}{\emph{Rev. Mod. Phys.} {\bfseries 81} (2009) 1199} [\href{https://arxiv.org/abs/0801.1345}{{\ttfamily 0801.1345}}].

\bibitem{Bondarenko:2018ptm}
K.~Bondarenko, A.~Boyarsky, D.~Gorbunov and O.~Ruchayskiy, \emph{{Phenomenology of GeV-scale Heavy Neutral Leptons}}, \href{https://doi.org/10.1007/JHEP11(2018)032}{\emph{JHEP} {\bfseries 11} (2018) 032} [\href{https://arxiv.org/abs/1805.08567}{{\ttfamily 1805.08567}}].

\bibitem{Boyarsky:2018tvu}
A.~Boyarsky, M.~Drewes, T.~Lasserre, S.~Mertens and O.~Ruchayskiy, \emph{{Sterile neutrino Dark Matter}}, \href{https://doi.org/10.1016/j.ppnp.2018.07.004}{\emph{Prog. Part. Nucl. Phys.} {\bfseries 104} (2019) 1} [\href{https://arxiv.org/abs/1807.07938}{{\ttfamily 1807.07938}}].

\bibitem{Peccei:1977hh}
R.D.~Peccei and H.R.~Quinn, \emph{{CP Conservation in the Presence of Instantons}}, \href{https://doi.org/10.1103/PhysRevLett.38.1440}{\emph{Phys. Rev. Lett.} {\bfseries 38} (1977) 1440}.

\bibitem{Weinberg:1977ma}
S.~Weinberg, \emph{{A New Light Boson?}}, \href{https://doi.org/10.1103/PhysRevLett.40.223}{\emph{Phys. Rev. Lett.} {\bfseries 40} (1978) 223}.

\bibitem{Wilczek:1977pj}
F.~Wilczek, \emph{{Problem of Strong $P$ and $T$ Invariance in the Presence of Instantons}}, \href{https://doi.org/10.1103/PhysRevLett.40.279}{\emph{Phys. Rev. Lett.} {\bfseries 40} (1978) 279}.

\bibitem{Choi:2020rgn}
K.~Choi, S.H.~Im and C.S.~Shin, \emph{{Recent progress in physics of axions or axion-like particles}},  \href{https://arxiv.org/abs/2012.05029}{{\ttfamily 2012.05029}}.

\bibitem{Antoniadis:2009ze}
I.~Antoniadis, A.~Boyarsky, S.~Espahbodi, O.~Ruchayskiy and J.D.~Wells, \emph{{Anomaly driven signatures of new invisible physics at the Large Hadron Collider}}, \href{https://doi.org/10.1016/j.nuclphysb.2009.09.009}{\emph{Nucl. Phys. B} {\bfseries 824} (2010) 296} [\href{https://arxiv.org/abs/0901.0639}{{\ttfamily 0901.0639}}].

\bibitem{Dror:2017nsg}
J.A.~Dror, R.~Lasenby and M.~Pospelov, \emph{{Dark forces coupled to nonconserved currents}}, \href{https://doi.org/10.1103/PhysRevD.96.075036}{\emph{Phys. Rev. D} {\bfseries 96} (2017) 075036} [\href{https://arxiv.org/abs/1707.01503}{{\ttfamily 1707.01503}}].

\bibitem{Borysenkova:2021ydf}
Y.~Borysenkova, P.~Kashko, M.~Tsarenkova, K.~Bondarenko and V.~Gorkavenko, \emph{{Production of Chern\textendash{}Simons bosons in decays of mesons}}, \href{https://doi.org/10.1088/1361-6471/ac77a7}{\emph{J. Phys. G} {\bfseries 49} (2022) 085003} [\href{https://arxiv.org/abs/2110.14500}{{\ttfamily 2110.14500}}].

\bibitem{Alekhin:2016}
S.~Alekhin et~al., \emph{{A facility to Search for Hidden Particles at the CERN SPS: the SHiP physics case}}, \href{https://doi.org/10.1088/0034-4885/79/12/124201}{\emph{Rept. Prog. Phys.} {\bfseries 79} (2016) 124201} [\href{https://arxiv.org/abs/1504.04855}{{\ttfamily 1504.04855}}].

\bibitem{Curtin:2018mvb}
D.~Curtin et~al., \emph{{Long-Lived Particles at the Energy Frontier: The MATHUSLA Physics Case}}, \href{https://doi.org/10.1088/1361-6633/ab28d6}{\emph{Rept. Prog. Phys.} {\bfseries 82} (2019) 116201} [\href{https://arxiv.org/abs/1806.07396}{{\ttfamily 1806.07396}}].

\bibitem{Asaka:2005}
T.~Asaka and M.~Shaposhnikov, \emph{{The $\nu$MSM, dark matter and baryon asymmetry of the universe}}, \href{https://doi.org/10.1016/j.physletb.2005.06.020}{\emph{Physics Letters B} {\bfseries 620} (2005) 17–26} [\href{https://arxiv.org/abs/hep-ph/0505013}{{\ttfamily hep-ph/0505013}}].

\bibitem{Asaka_2:2005}
T.~Asaka, S.~Blanchet and M.~Shaposhnikov, \emph{{The nuMSM, dark matter and neutrino masses}}, \href{https://doi.org/10.1016/j.physletb.2005.09.070}{\emph{Phys. Lett. B} {\bfseries 631} (2005) 151} [\href{https://arxiv.org/abs/hep-ph/0503065}{{\ttfamily hep-ph/0503065}}].

\bibitem{Akhmedov:1998qx}
E.K.~Akhmedov, V.A.~Rubakov and A.Y.~Smirnov, \emph{{Baryogenesis via neutrino oscillations}}, \href{https://doi.org/10.1103/PhysRevLett.81.1359}{\emph{Phys. Rev. Lett.} {\bfseries 81} (1998) 1359} [\href{https://arxiv.org/abs/hep-ph/9803255}{{\ttfamily hep-ph/9803255}}].

\bibitem{Buchmuller:2004nz}
W.~Buchmuller, P.~Di~Bari and M.~Plumacher, \emph{{Leptogenesis for pedestrians}}, \href{https://doi.org/10.1016/j.aop.2004.02.003}{\emph{Annals Phys.} {\bfseries 315} (2005) 305} [\href{https://arxiv.org/abs/hep-ph/0401240}{{\ttfamily hep-ph/0401240}}].

\bibitem{Pilaftsis:2005rv}
A.~Pilaftsis and T.E.J.~Underwood, \emph{{Electroweak-scale resonant leptogenesis}}, \href{https://doi.org/10.1103/PhysRevD.72.113001}{\emph{Phys. Rev. D} {\bfseries 72} (2005) 113001} [\href{https://arxiv.org/abs/hep-ph/0506107}{{\ttfamily hep-ph/0506107}}].

\bibitem{Davidson:2008bu}
S.~Davidson, E.~Nardi and Y.~Nir, \emph{{Leptogenesis}}, \href{https://doi.org/10.1016/j.physrep.2008.06.002}{\emph{Phys. Rept.} {\bfseries 466} (2008) 105} [\href{https://arxiv.org/abs/0802.2962}{{\ttfamily 0802.2962}}].

\bibitem{Pilaftsis:2009pk}
A.~Pilaftsis, \emph{{The Little Review on Leptogenesis}}, \href{https://doi.org/10.1088/1742-6596/171/1/012017}{\emph{J. Phys. Conf. Ser.} {\bfseries 171} (2009) 012017} [\href{https://arxiv.org/abs/0904.1182}{{\ttfamily 0904.1182}}].

\bibitem{Shaposhnikov:2009zzb}
M.~Shaposhnikov, \emph{{Baryogenesis}}, \href{https://doi.org/10.1088/1742-6596/171/1/012005}{\emph{J. Phys. Conf. Ser.} {\bfseries 171} (2009) 012005}.

\bibitem{Bodeker:2020ghk}
D.~Bodeker and W.~Buchmuller, \emph{{Baryogenesis from the weak scale to the grand unification scale}}, \href{https://doi.org/10.1103/RevModPhys.93.035004}{\emph{Rev. Mod. Phys.} {\bfseries 93} (2021) 035004} [\href{https://arxiv.org/abs/2009.07294}{{\ttfamily 2009.07294}}].

\bibitem{Klaric:2020phc}
J.~Klari\'c, M.~Shaposhnikov and I.~Timiryasov, \emph{{Uniting Low-Scale Leptogenesis Mechanisms}}, \href{https://doi.org/10.1103/PhysRevLett.127.111802}{\emph{Phys. Rev. Lett.} {\bfseries 127} (2021) 111802} [\href{https://arxiv.org/abs/2008.13771}{{\ttfamily 2008.13771}}].

\bibitem{Klaric:2021cpi}
J.~Klari\'c, M.~Shaposhnikov and I.~Timiryasov, \emph{{Reconciling resonant leptogenesis and baryogenesis via neutrino oscillations}}, \href{https://doi.org/10.1103/PhysRevD.104.055010}{\emph{Phys. Rev. D} {\bfseries 104} (2021) 055010} [\href{https://arxiv.org/abs/2103.16545}{{\ttfamily 2103.16545}}].

\bibitem{Drewes:2021nqr}
M.~Drewes, Y.~Georis and J.~Klaric, \emph{{Mapping the Viable Parameter Space for Testable Leptogenesis}}, \href{https://doi.org/10.1103/PhysRevLett.128.051801}{\emph{Phys. Rev. Lett.} {\bfseries 128} (2022) 051801} [\href{https://arxiv.org/abs/2106.16226}{{\ttfamily 2106.16226}}].

\bibitem{Drewes:2012ma}
M.~Drewes and B.~Garbrecht, \emph{{Leptogenesis from a GeV Seesaw without Mass Degeneracy}}, \href{https://doi.org/10.1007/JHEP03(2013)096}{\emph{JHEP} {\bfseries 03} (2013) 096} [\href{https://arxiv.org/abs/1206.5537}{{\ttfamily 1206.5537}}].

\bibitem{Buchmuller:1985jz}
W.~Buchmuller and D.~Wyler, \emph{{Effective Lagrangian Analysis of New Interactions and Flavor Conservation}}, \href{https://doi.org/10.1016/0550-3213(86)90262-2}{\emph{Nucl. Phys. B} {\bfseries 268} (1986) 621}.

\bibitem{Georgi:1991ch}
H.~Georgi, \emph{{On-shell effective field theory}}, \href{https://doi.org/10.1016/0550-3213(91)90244-R}{\emph{Nucl. Phys. B} {\bfseries 361} (1991) 339}.

\bibitem{Georgi:1992xg}
H.~Georgi, \emph{{Thoughts on effective field theory}}, \href{https://doi.org/10.1016/0920-5632(92)90003-B}{\emph{Nucl. Phys. B Proc. Suppl.} {\bfseries 29BC} (1992) 1}.

\bibitem{Grzadkowski:2010es}
B.~Grzadkowski, M.~Iskrzynski, M.~Misiak and J.~Rosiek, \emph{{Dimension-Six Terms in the Standard Model Lagrangian}}, \href{https://doi.org/10.1007/JHEP10(2010)085}{\emph{JHEP} {\bfseries 10} (2010) 085} [\href{https://arxiv.org/abs/1008.4884}{{\ttfamily 1008.4884}}].

\bibitem{Henning:2016lyp}
B.~Henning, X.~Lu and H.~Murayama, \emph{{One-loop Matching and Running with Covariant Derivative Expansion}}, \href{https://doi.org/10.1007/JHEP01(2018)123}{\emph{JHEP} {\bfseries 01} (2018) 123} [\href{https://arxiv.org/abs/1604.01019}{{\ttfamily 1604.01019}}].

\bibitem{Brivio:2017vri}
I.~Brivio and M.~Trott, \emph{{The Standard Model as an Effective Field Theory}}, \href{https://doi.org/10.1016/j.physrep.2018.11.002}{\emph{Phys. Rept.} {\bfseries 793} (2019) 1} [\href{https://arxiv.org/abs/1706.08945}{{\ttfamily 1706.08945}}].

\bibitem{DeVries:2020jbs}
J.~De~Vries, H.K.~Dreiner, J.Y.~G\"unther, Z.S.~Wang and G.~Zhou, \emph{{Long-lived Sterile Neutrinos at the LHC in Effective Field Theory}}, \href{https://doi.org/10.1007/JHEP03(2021)148}{\emph{JHEP} {\bfseries 03} (2021) 148} [\href{https://arxiv.org/abs/2010.07305}{{\ttfamily 2010.07305}}].

\bibitem{Broncano:2002rw}
A.~Broncano, M.B.~Gavela and E.E.~Jenkins, \emph{{The Effective Lagrangian for the seesaw model of neutrino mass and leptogenesis}}, \href{https://doi.org/10.1016/S0370-2693(02)03130-1}{\emph{Phys. Lett. B} {\bfseries 552} (2003) 177} [\href{https://arxiv.org/abs/hep-ph/0210271}{{\ttfamily hep-ph/0210271}}].

\bibitem{Coy:2018bxr}
R.~Coy and M.~Frigerio, \emph{{Effective approach to lepton observables: the seesaw case}}, \href{https://doi.org/10.1103/PhysRevD.99.095040}{\emph{Phys. Rev. D} {\bfseries 99} (2019) 095040} [\href{https://arxiv.org/abs/1812.03165}{{\ttfamily 1812.03165}}].

\bibitem{Broncano:2003fq}
A.~Broncano, M.B.~Gavela and E.E.~Jenkins, \emph{{Neutrino physics in the seesaw model}}, \href{https://doi.org/10.1016/j.nuclphysb.2003.09.011}{\emph{Nucl. Phys. B} {\bfseries 672} (2003) 163} [\href{https://arxiv.org/abs/hep-ph/0307058}{{\ttfamily hep-ph/0307058}}].

\bibitem{Jenkins:2013zja}
E.E.~Jenkins, A.V.~Manohar and M.~Trott, \emph{{Renormalization Group Evolution of the Standard Model Dimension Six Operators I: Formalism and lambda Dependence}}, \href{https://doi.org/10.1007/JHEP10(2013)087}{\emph{JHEP} {\bfseries 10} (2013) 087} [\href{https://arxiv.org/abs/1308.2627}{{\ttfamily 1308.2627}}].

\bibitem{Jenkins:2013wua}
E.E.~Jenkins, A.V.~Manohar and M.~Trott, \emph{{Renormalization Group Evolution of the Standard Model Dimension Six Operators II: Yukawa Dependence}}, \href{https://doi.org/10.1007/JHEP01(2014)035}{\emph{JHEP} {\bfseries 01} (2014) 035} [\href{https://arxiv.org/abs/1310.4838}{{\ttfamily 1310.4838}}].

\bibitem{Alonso:2013hga}
R.~Alonso, E.E.~Jenkins, A.V.~Manohar and M.~Trott, \emph{{Renormalization Group Evolution of the Standard Model Dimension Six Operators III: Gauge Coupling Dependence and Phenomenology}}, \href{https://doi.org/10.1007/JHEP04(2014)159}{\emph{JHEP} {\bfseries 04} (2014) 159} [\href{https://arxiv.org/abs/1312.2014}{{\ttfamily 1312.2014}}].

\bibitem{Ohlsson:2022hfl}
T.~Ohlsson and M.~Pernow, \emph{{One-loop matching conditions in neutrino effective theory}}, \href{https://doi.org/10.1016/j.nuclphysb.2022.115729}{\emph{Nucl. Phys. B} {\bfseries 978} (2022) 115729} [\href{https://arxiv.org/abs/2201.00840}{{\ttfamily 2201.00840}}].

\bibitem{Wang:2023bdw}
Y.~Wang, D.~Zhang and S.~Zhou, \emph{{Complete one-loop renormalization-group equations in the seesaw effective field theories}}, \href{https://doi.org/10.1007/JHEP05(2023)044}{\emph{JHEP} {\bfseries 05} (2023) 044} [\href{https://arxiv.org/abs/2302.08140}{{\ttfamily 2302.08140}}].

\bibitem{Zhang:2023kvw}
D.~Zhang, \emph{{Renormalization group equations for the SMEFT operators up to dimension seven}}, \href{https://doi.org/10.1007/JHEP10(2023)148}{\emph{JHEP} {\bfseries 10} (2023) 148} [\href{https://arxiv.org/abs/2306.03008}{{\ttfamily 2306.03008}}].

\bibitem{Zhang:2023ndw}
D.~Zhang, \emph{{Revisiting renormalization group equations of the SMEFT dimension-seven operators}}, \href{https://doi.org/10.1007/JHEP02(2024)133}{\emph{JHEP} {\bfseries 02} (2024) 133} [\href{https://arxiv.org/abs/2310.11055}{{\ttfamily 2310.11055}}].

\bibitem{Zhang:2021jdf}
D.~Zhang and S.~Zhou, \emph{{Complete one-loop matching of the type-I seesaw model onto the Standard Model effective field theory}}, \href{https://doi.org/10.1007/JHEP09(2021)163}{\emph{JHEP} {\bfseries 09} (2021) 163} [\href{https://arxiv.org/abs/2107.12133}{{\ttfamily 2107.12133}}].

\bibitem{Coy:2021hyr}
R.~Coy and M.~Frigerio, \emph{{Effective comparison of neutrino-mass models}}, \href{https://doi.org/10.1103/PhysRevD.105.115041}{\emph{Phys. Rev. D} {\bfseries 105} (2022) 115041} [\href{https://arxiv.org/abs/2110.09126}{{\ttfamily 2110.09126}}].

\bibitem{Du:2022vso}
Y.~Du, X.-X.~Li and J.-H.~Yu, \emph{{Neutrino seesaw models at one-loop matching: discrimination by effective operators}}, \href{https://doi.org/10.1007/JHEP09(2022)207}{\emph{JHEP} {\bfseries 09} (2022) 207} [\href{https://arxiv.org/abs/2201.04646}{{\ttfamily 2201.04646}}].

\bibitem{Fernandez-Martinez:2016lgt}
E.~Fernandez-Martinez, J.~Hernandez-Garcia and J.~Lopez-Pavon, \emph{{Global constraints on heavy neutrino mixing}}, \href{https://doi.org/10.1007/JHEP08(2016)033}{\emph{JHEP} {\bfseries 08} (2016) 033} [\href{https://arxiv.org/abs/1605.08774}{{\ttfamily 1605.08774}}].

\bibitem{Blennow:2023mqx}
M.~Blennow, E.~Fern\'andez-Mart\'\i{}nez, J.~Hern\'andez-Garc\'\i{}a, J.~L\'opez-Pav\'on, X.~Marcano and D.~Naredo-Tuero, \emph{{Bounds on lepton non-unitarity and heavy neutrino mixing}}, \href{https://doi.org/10.1007/JHEP08(2023)030}{\emph{JHEP} {\bfseries 08} (2023) 030} [\href{https://arxiv.org/abs/2306.01040}{{\ttfamily 2306.01040}}].

\bibitem{Ruchayskiy:2022}
K.A.~Urqu\'\i{}a-Calder\'on, I.~Timiryasov and O.~Ruchayskiy, \emph{{Heavy neutral leptons \textemdash{} Advancing into the PeV domain}}, \href{https://doi.org/10.1007/JHEP08(2023)167}{\emph{JHEP} {\bfseries 08} (2023) 167} [\href{https://arxiv.org/abs/2206.04540}{{\ttfamily 2206.04540}}].

\bibitem{Granelli:2022eru}
A.~Granelli, J.~Klari\'c and S.T.~Petcov, \emph{{Tests of low-scale leptogenesis in charged lepton flavour violation experiments}}, \href{https://doi.org/10.1016/j.physletb.2022.137643}{\emph{Phys. Lett. B} {\bfseries 837} (2023) 137643} [\href{https://arxiv.org/abs/2206.04342}{{\ttfamily 2206.04342}}].

\bibitem{Gorkavenko:2009vd}
V.M.~Gorkavenko and S.I.~Vilchynskiy, \emph{{Some constraints on the Yukawa parameters in the neutrino modification of the Standard Model (nuMSM) and CP-violation}}, \href{https://doi.org/10.1140/epjc/s10052-010-1488-y}{\emph{Eur. Phys. J. C} {\bfseries 70} (2010) 1091} [\href{https://arxiv.org/abs/0907.4484}{{\ttfamily 0907.4484}}].

\bibitem{Casas:2001sr}
J.A.~Casas and A.~Ibarra, \emph{{Oscillating neutrinos and $\mu \to e, \gamma$}}, \href{https://doi.org/10.1016/S0550-3213(01)00475-8}{\emph{Nucl. Phys.} {\bfseries B618} (2001) 171} [\href{https://arxiv.org/abs/hep-ph/0103065}{{\ttfamily hep-ph/0103065}}].

\bibitem{Tastet:2021vwp}
J.-L.~Tastet, O.~Ruchayskiy and I.~Timiryasov, \emph{{Reinterpreting the ATLAS bounds on heavy neutral leptons in a realistic neutrino oscillation model}}, \href{https://doi.org/10.1007/JHEP12(2021)182}{\emph{JHEP} {\bfseries 12} (2021) 182} [\href{https://arxiv.org/abs/2107.12980}{{\ttfamily 2107.12980}}].

\bibitem{Abada:2006ea}
A.~Abada, S.~Davidson, A.~Ibarra, F.X.~Josse-Michaux, M.~Losada and A.~Riotto, \emph{{Flavour Matters in Leptogenesis}}, \href{https://doi.org/10.1088/1126-6708/2006/09/010}{\emph{JHEP} {\bfseries 09} (2006) 010} [\href{https://arxiv.org/abs/hep-ph/0605281}{{\ttfamily hep-ph/0605281}}].

\bibitem{ParticleDataGroup:2022pth}
{\scshape Particle Data Group} collaboration, \emph{{Review of Particle Physics}}, \href{https://doi.org/10.1093/ptep/ptac097}{\emph{PTEP} {\bfseries 2022} (2022) 083C01}.

\bibitem{ALEPH:2005ab}
{\scshape ALEPH, DELPHI, L3, OPAL, SLD, LEP Electroweak Working Group, SLD Electroweak Group, SLD Heavy Flavour Group} collaboration, \emph{{Precision electroweak measurements on the $Z$ resonance}}, \href{https://doi.org/10.1016/j.physrep.2005.12.006}{\emph{Phys. Rept.} {\bfseries 427} (2006) 257} [\href{https://arxiv.org/abs/hep-ex/0509008}{{\ttfamily hep-ex/0509008}}].

\bibitem{Pich:2013lsa}
A.~Pich, \emph{{Precision Tau Physics}}, \href{https://doi.org/10.1016/j.ppnp.2013.11.002}{\emph{Prog. Part. Nucl. Phys.} {\bfseries 75} (2014) 41} [\href{https://arxiv.org/abs/1310.7922}{{\ttfamily 1310.7922}}].

\end{thebibliography}\endgroup

\end{document}